\documentclass[12pt,preprint]{aastex}

\shorttitle{}
\shortauthors{Luo et al.}

\begin{document}

\title{Hot subdwarf stars observed in LAMOST DR1 - Atmospheric parameters from single-lined spectra}

\author{Yang-Ping Luo\altaffilmark{1,2,3},P$\acute{\rm{e}}$ter N$\acute{\rm{e}}$meth\altaffilmark{4}, Chao  Liu\altaffilmark{1},  Li-Cai Deng\altaffilmark{1} and Zhan-Wen Han\altaffilmark{3} }

\altaffiltext{1}{Key Laboratory of Optical Astronomy, National Astronomical Observatories, Chinese Academy of Sciences, Beijing 100012, China \email{ypluo@bao.ac.cn}}
\altaffiltext{2}{School of Physics and Space Science, China West Normal University, Nanchong, 637009, China}
\altaffiltext{3}{Key Laboratory for the Structure and Evolution of Celestial Objects, Chinese Academy of Sciences, Kunming 650011, China}
\altaffiltext{4}{Dr. Remeis-Sternwarte Astronomisches Institut,  Universit$\ddot{\rm{a}}$t  Erlangen-N$\ddot{\rm{u}}$rnberg , 96049, Bamberg, Germany}

\begin{abstract}
We present a catalog of 166 spectroscopically identified hot subdwarf stars from LAMOST DR1, 44 of which show the characteristics of cool companions
in their optical spectra. Atmospheric parameters of 122 non-composite spectra subdwarf stars were measured by fitting
the profiles of hydrogen (H) and helium (He) lines with synthetic spectra from non-LTE model atmospheres.
Most of the sdB stars scatter near the Extreme Horizontal Branch in the $T_{\rm eff}-\log{g}$ diagram and two well defined groups can be outlined.
A clustering of He-enriched sdO stars appears near $T_{\rm eff}=45\,000$ K and $\log(g) = 5.8$. The sdB population separates into several nearly parallel sequences in the $T_{\rm eff}-{\rm He}$ abundance diagram with clumps corresponding to those in the $T_{\rm eff}-\log{g}$ diagram.
Over $38\,000$ K (sdO) stars show abundance extremes, they are either He-rich or He-deficient and we observe only a few stars in the $ -1 < \log(y) < 0$ abundance range. With increasing temperature these extremes become less prominent and the He abundance approaches to $\log(y)\sim-0.5$.
A unique property of our sample is that it covers a large range in apparent magnitudes and galactic latitudes, therefore it contains a mix of stars from different populations and galactic environments. Our results are consistent with the findings of Hirsch (2009) and we
conclude that He-rich and He-deficient sdB stars ($\log(y) < 1$) probably origin from different populations. We also find that most sdO and sdB stars lie in a
narrow strip in the luminosity and helium abundance plane, which suggests that
these atmospheric parameters are correlated.

\end{abstract}

\section{Introduction}
Hot subdwarf stars are core He burning stars with a canonical mass of $M\sim0.5M_{\odot}$, having a very thin hydrogen envelope \citep{Heber2009}. They are located at the blue end of the Horizontal Branch (HB), also named as the Extreme Horizontal Branch (EHB) \citep{Heber1984} in the Hertzsprung-Russell diagram (HRD). In general, they can be classified as the cooler sdB stars, whose spectra typically show dominant H lines and weak He$\,$I lines, and the hotter sdO stars, which are characterized by He$\,$I lines and weak He$\,$II lines in their spectra and exhibit a higher He abundance on average \citep{Drilling2003, Stroeer2007, Heber2009}. There are other classes named as the He-sdB and He-sdO stars. They are different from sdB and sdO stars and have almost pure He atmospheres.
The origin of hot subdwarf stars is still unknown and subject of extensive research, largely because they are the main source of the UV-upturn in the spectra of elliptical galaxies and the bulge of spiral galaxies \citep{Connell1999, Han2007}. Hot subdwarfs are also important to understand the
horizontal branch morphology of globular clusters \citep{Han2008, Lei2013, Lei2015}.
They are also very important in stellar astrophysics.
The discovery of pulsating subdwarfs provided an excellent environment to probe their interior structure using the tools of asteroseismology \citep{Fontaine2012}. Moreover, they are even relevant for cosmology, as some of them may qualify as Supernova Ia progenitors \citep{Geier2007,Wang2010}.

A number of scenarios trying to explain the formation and evolution of hot subdwarf stars have been put forward.
In the canonical formation scenarios binary evolution \citep{Han2002, Han2003} and enhanced mass loss \citep{Han1994} from single red-giant stars are responsible for sdB stars, and double white dwarf (WD) mergers for He-rich sdO stars \citep{Webbink1984, Han2002, Justham2011, Zhang2012}.
Among non-standard formation scenarios the hot-flasher scenario \citep{DCruz1996,Sweigart1997,Lanz2004,Miller2008} can reproduce the observed abundance diversities.
Although these models can account for the observed properties of hot subdwarfs, none of them appears entirely satisfactory, largely because some key physical processes (mass loss on the red-giant branch, surface element diffusion, common-envelope evolution, mass transfer, etc.) are not dealt with satisfactorily \citep{Heber2009, Han2010, Nemeth2012}.
The currently available observations \citep{Edelmann2003, Lisker2005, Stroeer2007, Hirsch2009, Nemeth2012, Geier2013, Geier2013a,Geier2015} cannot provide enough information on
their origin and evolutionary status.
Therefore, spectral analyses on large and homogeneous samples are still very valuable because some new observational constraints can be outlined.

LAMOST (the Large Sky Area Multi-Object Fiber Spectroscopic Telescope, also named the Guo Shou Jing Telescope)
is a 4-m specially designed Schmidt survey telescope at the Xinglong Station of the National Astronomical Observatories of Chinese Academy of
Sciences, which can simultaneously take the spectra of 4000 objects in an about $5^{\circ}$ (diameter) field of view \citep{Cui2012}. It is equipped with 16 low-resolution spectrographs, 32 CCDs and 4,000 optical fibres.
The LAMOST survey set an objective to observe at least 2.5 million stars
in a contiguous area in the Galactic halo and more than 7.5 million stars at low galactic latitudes
within four years \citep{Zhao2012,Deng2012}. From October 2011 to June 2013, the LAMOST survey has obtained
more than 2 million spectra, which were released as the DR1 catalog, in which the signal-to-noise ratio (S/N) of about 1.2 million spectra is more than 10 \citep{Zhaoy2014, Liu2014}. Therefore, this huge spectral database provides an opportunity to search for hot subdwarfs and perform a spectral analysis on this large and homogeneous sample.

This paper, as our first work, reports 166 spectroscopically identified hot subdwarfs from LAMOST DR1 and presents a spectral analysis for 122 non-composite spectra targets. In Section 2, we describe the data and target selection. Section 3 contains a brief
description of the atmospheric models and the determination of atmospheric parameters. Our results and discussions are given in Sect$\,.$4 and a summary follows in Sect$\,.$5.

\section{LAMOST DR1 Data and Target Selection}
LAMOST DR1 has released more than 2 million spectra, including about 700,000 spectra from the pilot survey \citep{Luo2012c}.
There are 1.7 million spectra of stars, in which the stellar parameters (effective temperature, surface gravity, metallicity and radial velocity) of over 1 million stars were acquired \citep{Zhaoy2014}. The target selection algorithm, survey design and observations in the LAMOST Spectroscopic Survey have been presented in \citet{Carlin2012,Chenl2012,Zhangy2012a,Yangf2012,Xiang2015,Yuan2015}. The LAMOST spectra are similar to the SDSS data having a resolving power of $R \sim 1800$
and covering the wavelength range from $3800\rm{\AA}$  to $9100\rm{\AA}$.
Three data pipelines have been developed for the LAMOST survey \citep{Luo2012, Luo2014b}.
The raw spectra were reduced by using the standard LAMOST 2D pipeline, including bias subtraction, cosmic-ray
removal, spectral trace and extraction, flat-fielding, wavelength calibration, sky subtraction, and combination.
The classification and redshift measurement of the extracted spectra were done with the 1D pipeline and the stellar parameters were measured by the LASP (LAMOST Stellar Parameters) pipeline. And recently, another stellar parameter pipeline LSP3 \citep{Xiang2015} have been developed by another group at Peking University to compare with the LASP.

However, the classification of LAMOST spectra is not suitable for hot subdwarf stars, because they are not included in the stellar templates. Therefore, our selection of hot subdwarf candidates was done differently, in two ways:
First, we used the SDSS $ugr$ magnitudes to select the candidates from the LAMOST DR1 catalog. Hot stars are found easily by colour cuts on SDSS photometry \citep{Geier2011}.
Our initial 320,734 spectra with SDSS $ugr$ magnitudes were selected from the LAMOST DR1 catalog,
in which the $S/N$ of 115,791 spectra is more than 10.
Next, we obtained 462 candidates by taking advantage of the colour cuts $[-0.6<u-g<0.4\,\& -0.7<g-r<0.1]$ defined by  \citet{Geier2011} on the basis of a hot subdwarf sample from UV excess surveys \citep{Green1986, Jester2005}.
As described in \cite{Geier2011}, the color criteria ensures that sdB spectroscopic binaries with dwarf companions of spectral type $F$ or later are
included while the huge numbers of QSOs (quasi stellar objects) are not. After rejecting bad spectra and other targets (QSOs, white dwarfs, main sequence stars), we obtained 74 stars by visual comparisons with reference spectra of hot subdwarfs.
Second, we cross-correlated our sample with the archived database of hot subdwarf candidates.
We collected 3,868 archived hot subdwarf candidates from the VizieR database \citep{Ochsenbein2000} and from the Hot Subdwarf Database \citep{Ostensen2004},
196 of which we also found in the LAMOST DR1 catalog. The resulting 145 stars have good spectra ($S/N>10$) making them suitable for a spectral analysis.

\begin{figure*}
\epsscale{0.6}
\plotone{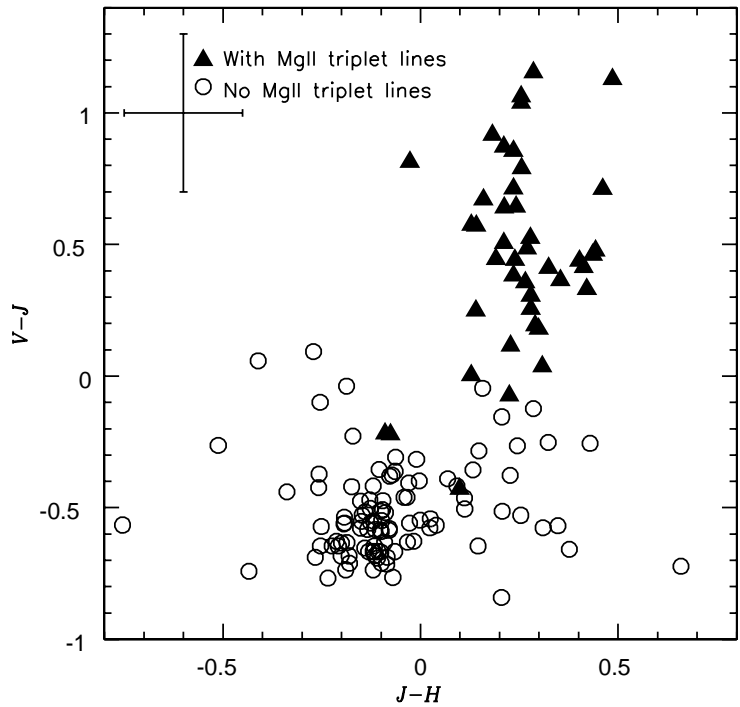}
\caption{ Two-color plot of $V-J$ versus $J-H$ for 148 hot subdwarf stars in LAMOST DR1. The triangles denote the spectra with MgII triplet lines and the circles represent the spectra without MgII triplet lines. \label{fig1}}
\end{figure*}

By combining the two parts, a final sample of 166 hot subdwarf stars has been obtained from LAMOST DR1, in which 44 stars show strong double-line composite spectra. They show noticeable MgII triplet lines at 5170 $\rm{\AA}$ or CaII triplet lines at 8650 $\rm{\AA}$, which were taken as indications of a late type companion \citep{Heber2009}.  But the latter are seriously polluted by sky emission lines in LAMOST spectra.
Their parameters are listed in Table$\,$\ref{tbl1-1}.
Figure$\,$\ref{fig1} shows a two-color diagram of $V-J$ versus $J-H$  for only 148 hot subdwarf stars in our sample because these two colors are no available for other 18 ones. Optical $V$ magnitudes were collected from GSC2.3.2 \citep{Lasker2008} and  UCAC4 \citep{Zacharias2013} and infrared (IR) $JH$ magnitudes from 2MASS \citep{Cutri2003}.  At least 22 per cent of our sample has $V-J>0$ and $J-H>0$ and shows IR excess. This number is close to 19 per cent in GALEX sample \citep{Nemeth2012}.

\section{Atmospheric parameters}
We applied the non-LTE model atmosphere code {\sc Tlusty} \citep{hubeny95} and spectral synthesis code {\sc Synspec} \citep{lanz07} to calculate subdwarf spectral models with H-He composition. {\sc Tlusty} calculates model atmospheres in hydrostatic and radiative equilibrium in plane-parallel geometry. Atomic data were taken from the {\sc Tlusty} website and Stark broadening data for the hydrogen lines from \cite{lemke97} and \citet{tremblay11}.

\begin{figure*}
\epsscale{1.0}
\plotone{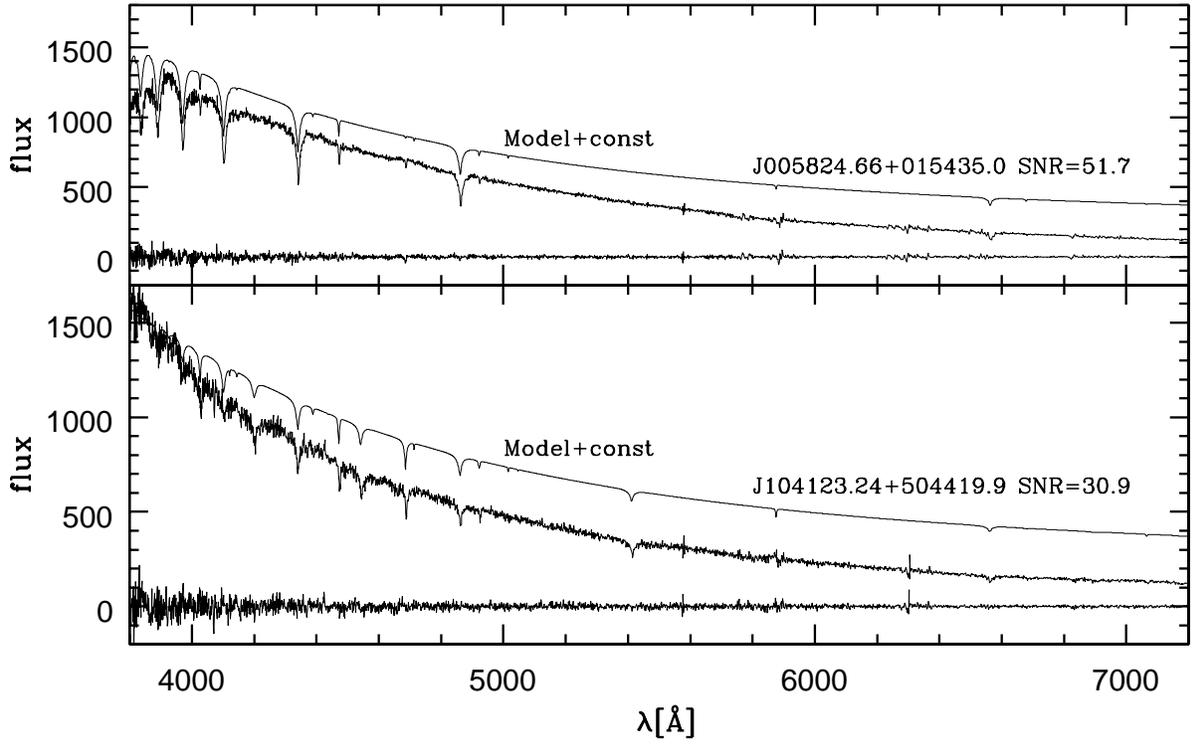}
\caption{Example fits of the observed spectra with model atmospheres for two hot subdwarfs in our fitting range.
Top: $T_{\rm{eff}}=32350\pm450\rm{K}$, $\log(g)=5.713\pm0.118$, $\log(y)=-1.925\pm0.114$. Bottom:$T_{\rm{eff}}=51720\pm1690\rm{K}$, $\log(g)=5.884\pm0.199$, $\log(y)=0.588\pm0.569$. \label{fig2}}
\end{figure*}

Atmospheric parameters (effective temperature $T_{\rm eff}$, surface gravity $\log{g}$ and He abundances $y=n(\rm He)/n({\rm H}))$ were measured by fitting synthetic spectra, normalized in 80 \AA sections, to the flux calibrated observations. We applied our steepest-decent chi-square minimization spectral analysis procedure ({\sc XTgrid}; \cite{Nemeth2012}) to fit the sample. {\sc XTgrid} was designed to work with {\sc Tlusty} models and perform a fully automatic parameter determination for large samples. The procedure calculates new models in the direction of decreasing chi-squares, therefore it does not require a grid and seamlessly covers the transition between the sdO and sdB spectral types, where spectra show a great diversity. We used the $3800 - 7200$ \AA\, range that includes all the significant H and He lines in the LAMOST spectra. Two example are shown in Figure$\,$\ref{fig2}. The data range was limited in a small subset of the spectra to avoid artifacts changing the results. Although some spectra show metal lines, in particular C, N, Mg and Si lines, the SNR in general does not allow for a detailed abundance analysis.

We analyzed only the 122 non-composite spectra stars in this paper and leave the composite spectra for a forthcoming work. The median parameter errors are: $T_{\rm eff}=1030$ K, $\log{g} = 0.16$ cm s$^{-2}$ and $\log{y} = 0.29$ dex, although we note, that the error bars show a strong correlation with spectral types and the SNR of the data.

\subsection{ Overlaps with other catalogs }

\subsubsection{ GALEX Survey }

Two of our targets have been reported in the low-resolution survey of hot subluminous stars in the GALEX survey \citep{Nemeth2012}, which used the same analysis procedure.
For LAMOST\,J011928.87+490109.3 they found
$T_{\rm eff}=43720^{+510}_{-500}$ K,
$\log{g}=5.86^{+0.07}_{-0.21}$ cm$\,s^{-2}$ and
helium abundance $\log(y)=0.158^{+0.442}_{-0.048}$
in agreement with the current parameters:
$T_{\rm eff}=42660\pm780$ K,
$\log{g}=5.84\pm0.21$ cm$\,s^{-2}$ and
helium abundance $\log(y)=0.295\pm0.315$.
For LAMOST\,J085649.26+170114.6 they found
$T_{\rm eff}=29270^{+380}_{-450}$ K,
$\log{g}=5.39^{+0.20}_{-0.03}$ cm$\,s^{-2}$ and
an upper limit on the helium abundance $\log(y)<-2.81$
in agreement with our parameters:
$T_{\rm eff}=29360\pm230$ K,
$\log{g}=5.48\pm0.06$ cm$\,s^{-2}$ and
helium abundance $\log(y)=-3.101\pm0.199$.
These numbers show a reassuring internal consistency between the two analyses.

\subsubsection{SDSS Survey and PG Survey }
\begin{figure*}
\epsscale{0.6}
\plotone{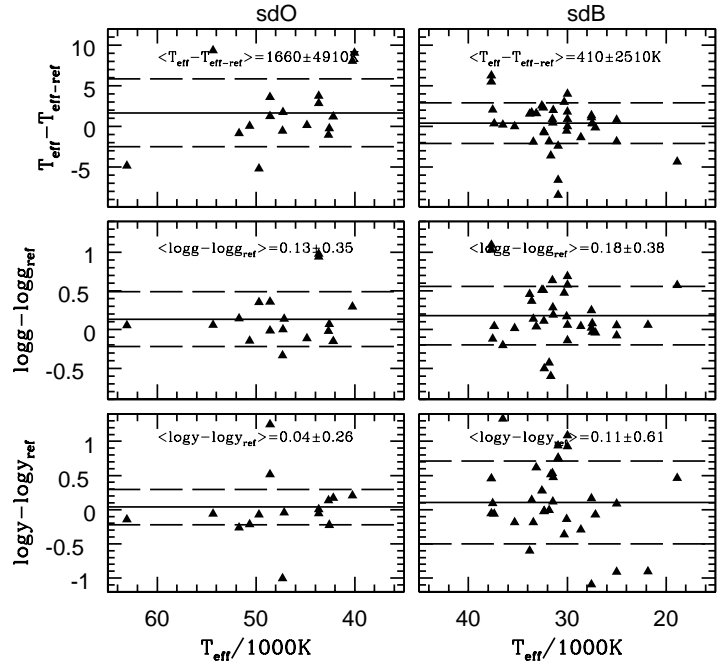}
\caption{Comparison with the literatures for $T_{\rm{eff}}$, $\log(g)$, $\log(y)$. The solid lines represent the averages of the shifts with respect to literature results and the dashed lines denote $1\sigma$ fitting error.\label{fig3}}
\end{figure*}

Many of our stars have been observed in the Palomar-Green Survey \citep{pg1986} and  the Sloan Digital Sky Survey (SDSS) and their spectra have been analyzed with a variety of methods. In Table \ref{tbl2-2} we list the identifications of these targets with references to past works. To find systematic effects we collected atmospheric parameters on these stars and calculated the differences with our parameters in Figure \ref{fig3}. We found the mean shifts in $T_{\rm eff}$, $\log{g}$ and He abundance $\log(y)$ are $\Delta(T_{\rm eff})=1660\pm4910$ K, $\Delta(\log{g})=0.13\pm0.35$, $\Delta(\log{y})=0.04\pm0.26$ for sdO stars, and $\Delta(T_{\rm eff})=410\pm2510$ K, $\Delta(\log{g})=0.18\pm0.38$, $\Delta(\log{y})=0.11\pm0.61$ for sdB stars.
These numbers show that our sample is comparable to published results. Such systematic shifts are quite general when parameters from different model atmosphere codes are compared. Our results are based non-LTE model atmospheres with H+He composition, while the majority of the PG sample were analized  with metal line blanketed LTE models. The Stark line-broadening tables we used also change $T_{\rm eff}$ and $\log{g}$ upward.

\section{Results and Discussions}
Table$\,$\ref{tbl2-2} summarizes the results of our analysis of 122 non-composite stars, including the effective temperatures $T_{\rm{eff}}$, surface gravities $\log(g)$ and He abundances $(y=n(\rm{He})/\textit{n}(\rm{H}))$. Our spectral classification follows the scheme of \cite{Nemeth2012}.
Out of the 122 stars we identified 27 sdO and 88 sdB stars.
We discuss the subdwarf formation channels and evolutionary status based on the statistical properties of the LAMOST sample in the $T_{\rm{eff}}-\log(g)$, $T_{\rm{eff}}-\log(y)$, and $\log(L/L_{\rm{edd}})-\log(y)$ planes. In order to clearly display the statistical properties, sdB stars are grouped into He-rich and He-deficient ones by using the solar He abundance $\log(y)=-1$ \citep{Edelmann2003}. Similarly, sdO stars are divided into He-rich and He-deficient groups.

Checking the completeness of the sample is very important before looking for the statistical properties and comparing them to the predictions of theoretical models. However, it is ignored in this paper because our sample suffers from some uncertain selection effects. Hot subdwarf stars are not the primary science targets of the LAMOST survey and the target selection was based on different catalogs by using the different methods \citep{Carlin2012,Chenl2012,Zhangy2012a} due to the lack of a homogeneous multi-color photometric survey for the LAMOST sky area, unlike for SDSS. The different observational strategies were discussed by \cite{Yangf2012} and \cite{Xiang2015}. In general, the number ratio between sdB and sdO stars have been found to be around 3 from the previous surveys \citep{Heber2009, Ostensen2004, Nemeth2012}. However, this ratio in our sample is about 5, which shows that our sample suffers from selection effects, in particular for sdO stars. Therefore, these selection effects should be taken into account in the following discussions.
Although the target selection of the LAMOST surveys does not allow us to do any statistics yet, the sample studied here is just the tip of the iceberg and the combination with upcoming and present photometric surveys (e.g. UVEX, IPHAS, VPHAS+, PanSTARRS, VST-ATLAS, Skymapper, etc.) might substantially increase the number of sdO and sdB stars found in LAMOST survey in the future.

\subsection{Effective temperature and surface gravity}
\begin{figure*}
\epsscale{0.9}
\plotone{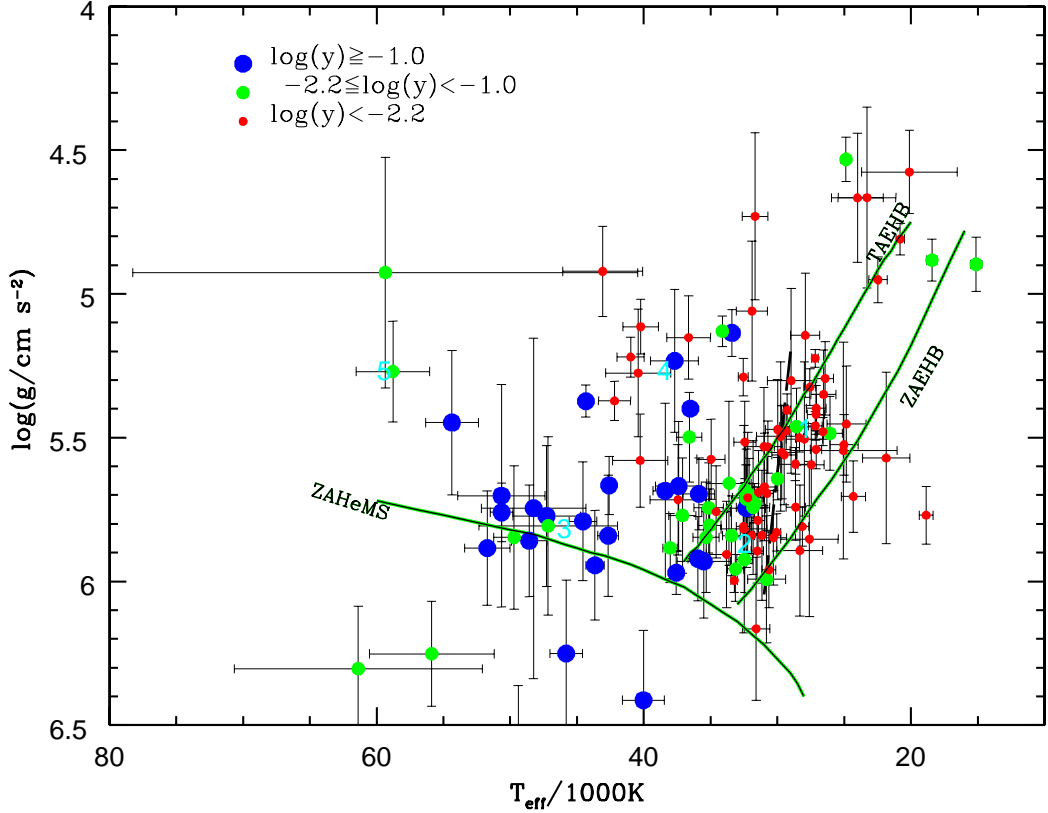}
\caption{$T_{\rm{eff}}-\log(g)$ diagram. The zero-age EHB (ZAEHB), terminal-age EHB (TAEHB) \citep{Dorman1993}, and zero-age He main sequence (ZAHeMS) \citep{paczynski1971} are marked with the green lines, respectively. The dashed line denotes the observed boundary of slow (Right) and rapid (Left) pulsating sdB stars \citep{Charpinet2010}.
\label{fig4}}
\end{figure*}

Figure$\,$\ref{fig4} displays the distribution of our sample in the $T_{\rm{eff}}-\log(g)$ plane. We also plot
the location of the EHB band as shown in Figure$\,5$ in \cite{Nemeth2012}, which is defined as the region between the zero age extended horizontal branch (ZAEHB) and the terminal age extended horizontal branch (TAEHB) derived from evolutionary tracks of \cite{Dorman1993} for solar metallicity. We also show the location of the zero age He main sequence (ZAHeMS) by \cite{paczynski1971} and the observed boundary of $g-$mode and $p-$mode pulsating sdB stars from \cite{Charpinet2010}.

 Most of the sdB stars in our sample lie in the EHB band. There is a known shift \citep{Nemeth2012} with respect to TAEHB for solar metallicity and it is noticeable in our sample as well. This may be due to that not enough metals are included in the non-LTE models \citep{hubeny95} or due to different He core masses \citep{Han2002}.
As seen in \cite{Nemeth2012}, sdB stars show two groups (no.1 and 2) on the EHB.
Group 1 is the cooler, He-poor sdB stars that crowd around $T_{\rm{eff}}=28\,000\,$K  and $\log(g)=5.4$.
They lie to the right of the observed boundary of $g-$mode and $p-$mode pulsating sdB stars and are potential $g-$mode pulsators.
Group 2 is the hotter sdB stars that are found around $T_{\rm{eff}}=33\,500\,$K and $\log(g)=5.8$ and on average ten times more He abundant than group 1. They are located to the left of the observed boundary of $g-$mode and $p-$mode pulsating sdB stars and possible $p-$mode pulsators.
We also see two groups (no.3 and 4) among the sdO stars.  One is the He-rich sdO group (Group 3) between $40\,000$ and $50\,000$K near $\log(g)=5.8$ around the theoretical HeMS, another (Group 4) is the mixture of He-deficient and He-rich sdO/B stars around $T_{\rm{eff}}=38000\,$K and $\log(g)=5.3$.
There are one He-rich star and two He-deficient sdO stars in region 5, but our sample is not large enough to outline any significant groups in this region.
These observations are in good agreement with the results reported by \cite{Hirsch2009} and \cite{Nemeth2012}.

Furthermore, He-deficient sdO stars are scattered in a wider range in the $T_{\rm{eff}}-\log(g)$ planes and no correlation can be detected.
The number ratio between He-rich and He-deficient sdO stars is around $1.4$ which is closer to $1.6$ in the sample of \cite{Nemeth2012}, but lower than $2.5$ in the sample of \cite{Stroeer2007} maybe due to selection effects.
Our sample supports that He-rich sdO stars are probably more frequent than He-deficient ones \citep{Heber2009}, in agreement with other samples \citep{Stroeer2007, Nemeth2012}. It is likely that these two classes of sdO stars origin from different formation channels, He-rich sdO stars are from the double WDs merger \citep{Zhang2012} and He-deficient ones from the evolution of the He-deficient sdB \citep{Dorman1993}.

\begin{figure*}
\epsscale{0.9}
\plotone{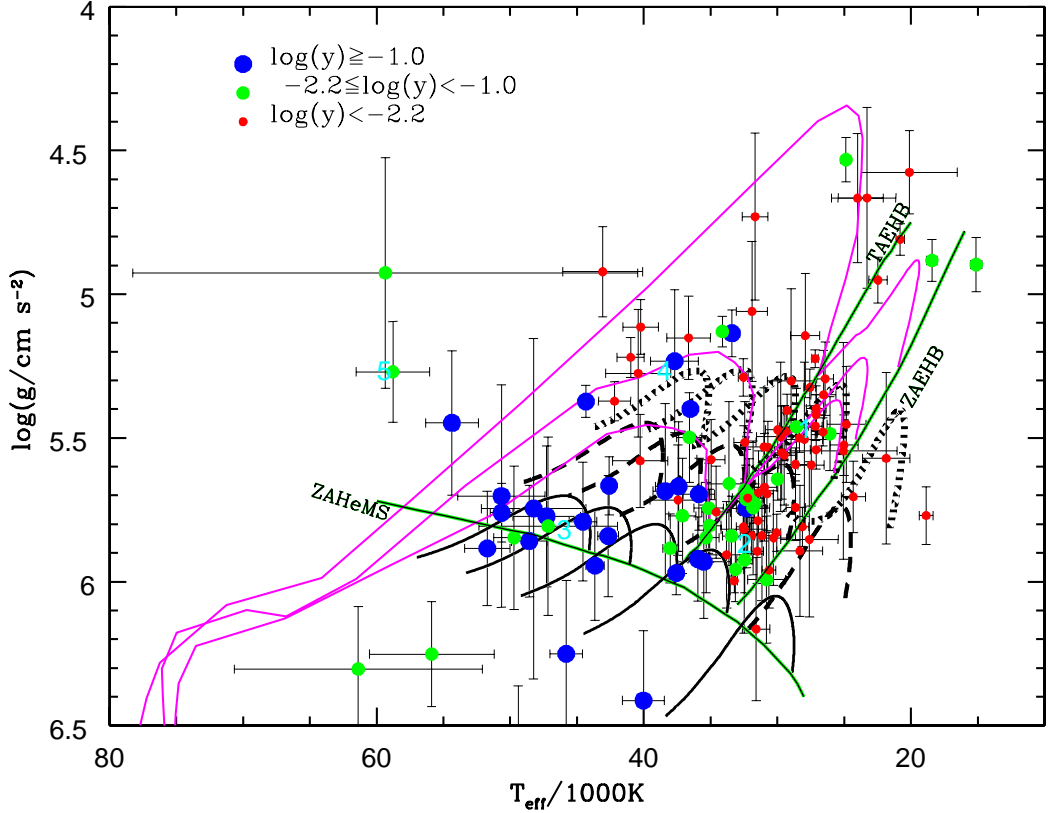}
\caption{$T_{\rm{eff}}-\log(g)$ diagram for the canonical formation scenario.  The magenta curves are the evolutionary tracks of
\cite{Dorman1993} for solar metallicity and subdwarf masses from top to bottom: 0.480, 0.473 and $0.471\,M_{\odot}$. The dark curves from right to left show the sdB evolutionary tracks of \cite{Han2002} for sdB masses of 0.35, 0.45,  0.55,  0.65,  and $0.75\,M_{\odot}$ from the ZAHB to the point of central He exhaustion. The dark solid curves are for an envelope mass of $0.000\,M_{\odot}$, the dark dashed curves for $0.002\,M_{\odot}$ and the dark dotted curves for $0.005\,M_{\odot}$.
\label{fig5}}
\end{figure*}

\begin{figure*}
\epsscale{0.9}
\plotone{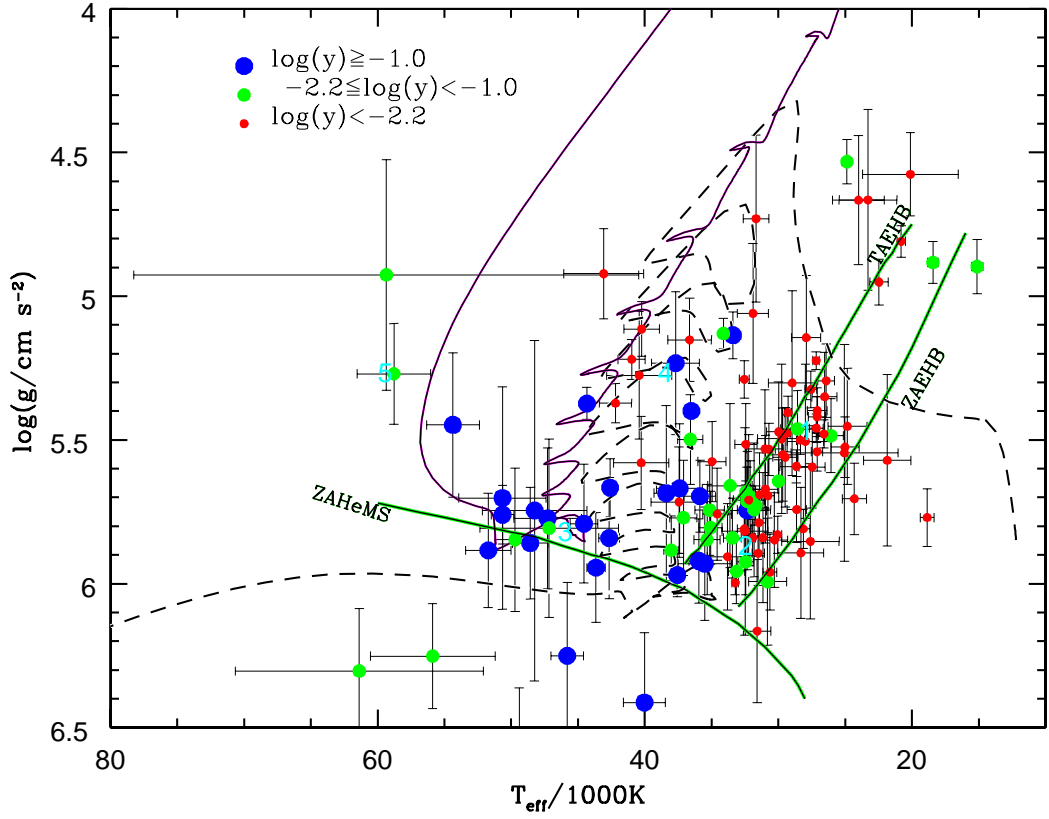}
\caption{$T_{\rm{eff}}-\log(g)$ diagram for the double WDs merger channels.
The solid and dashed curves denote the evolutionary tracks for subdwarf mass of 0.8 and $0.5\,M_{\odot}$ through the double WDs merger channels \citep{Zhang2012}.\label{fig6}}
\end{figure*}

To solve the puzzle of the origin and evolution of sdB and sdO stars and find potential links between both classes of stars, a number of scenarios have been put forward. The main scenarios are: the canonical EHB and post-EHB evolution \citep{Dorman1993}, canonical binary evolution \citep{Han2002,Zhang2012} and non-canonical hot-flasher scenario \citep{Miller2008}. To test these scenarios, we compare our observational results to their evolutionary tracks.
Figure \ref{fig5} shows subdwarf evolutionary tracks \citep{Dorman1993} from the EHB through the post-EHB phase for subdwarf masses of 0.471, 0.473 and 0.480$\,M_{\odot}$ and solar metallicity.
We can see that the post-EHB evolutionary tracks overlap with group 4 around $T_{\rm{eff}}=38000\,$K and $\log(g)=5.3$,
but they fail to explain this group. The calculations by \cite{Dorman1993} suggest that the evolutionary timescales are practically constant through the post-EHB phase and the lifetime is an order of magnitude shorter than on the EHB.
In Figure$\,$\ref{fig5}, we also mark the evolutionary tracks \citep{Han2002} for three sets of hot subdwarf stars with envelope masses of 0.000, 0.002 and $0.005\,M_{\odot}$ and a metallicity of $Z=0.02$ . For each set, we distinguish hot subdwarf masses of 0.35, 0.45, 0.55, 0.65, and $0.75\, M_{\odot}$. By comparing our sample to these evolutionary tracks, our sample could be explained by the hot subdwarf stars with different He core and envelope masses.
This is consistent with the prediction of the canonical model \citep{Han2002,Han2003}.
The canonical subdwarf formation theory \citep{Han2002} proposed three main formation channels: the common-envelope (CE) ejection channel, the stable Roch lobe overflow (RLOF) channel, and the double white dwarf (WDs) merger channel. They provide a good interpretation for the formation of sdB and sdO stars. Binary population synthesis models predict distinct properties of subdwarfs from the different channels. Therefore it is tempting to associate the lower temperature and surface gravity group 1 with the Common-Envelope formation channel, and the higher temperature and gravity group 2 with the Roche-lobe Overflow channel. \cite{Nemeth2012} found that long-period composite spectra binaries (sdB+F/G) from the Roche-lobe overflow channel show up exclusively in the higher temperature and gravity group. However, observations \citep{Kawka2015, Kupfer2015} show that both short- and long-period binaries occur in each group, suggesting that they have a mixture of stars with different formation history. The existence of the two sdB groups in the temperature gravity plane is an important result, but further investigations are needed to find their significance and weather we can infer from these groups to the yield of various formation channels \citep{Han2003}. Recently, \cite{Zhang2012} carried out extensive calculations for the double He WDs merger. Their evolutionary tracks for subdwarf masses of 0.5 and $0.8\,M_{\odot}$ and solar metallicity are presented in Figure$\,$\ref{fig6}, which shows that the He-rich sdO group 3 could also be explained well with the merger channel.

\begin{figure*}
\epsscale{0.9}
\plotone{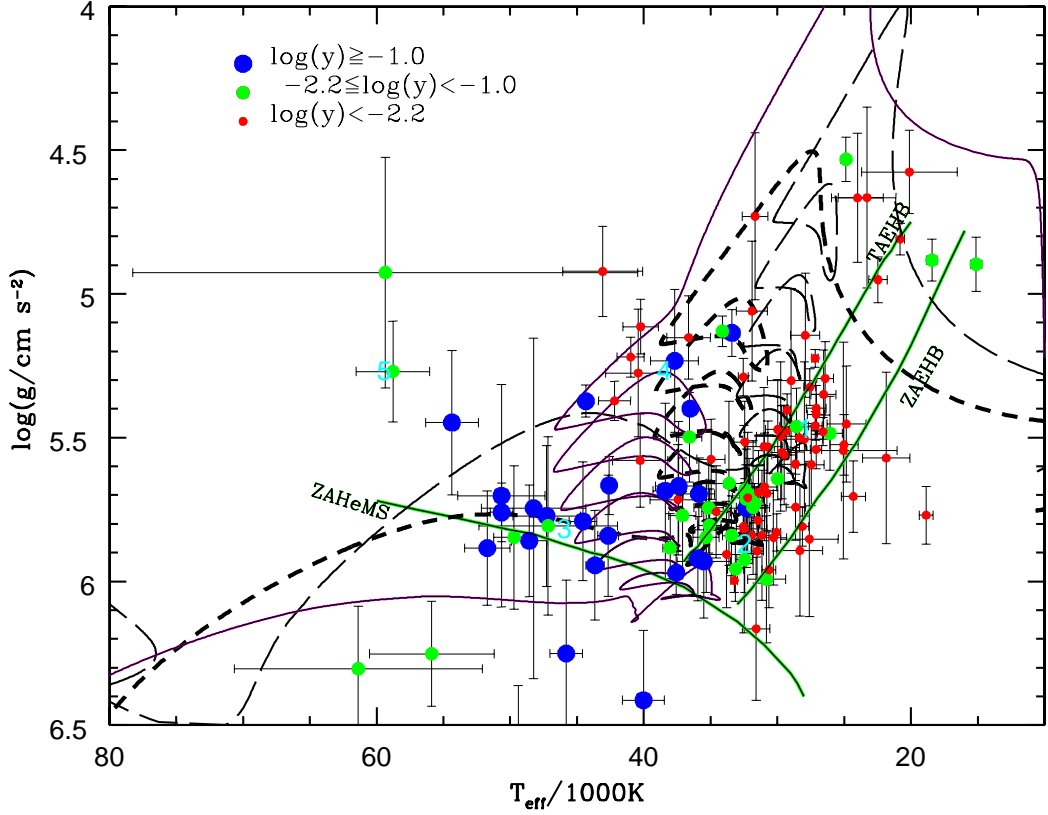}
\caption{$T_{\rm{eff}}-\log(g)$ diagram for the hot-flasher scenario.
The long dashed curve represents the evolutionary tracks for subdwarf mass of $0.47426\,M_{\odot}$ through the hot-flasher scenario with no He enrichment, the short dashed curve for subdwarf mass of $0.47378\,M_{\odot}$ with shallow mixing (SM), and the solid curve for subdwarf mass of $0.47112\,M_{\odot}$ with deep mixing (DM) \citep{Miller2008}.\label{fig7}}
\end{figure*}

Another comparison should be made with a non-canonical scenario named as hot-flasher \citep{Miller2008}. Its main feature is that stars experience a delayed core flash after the giant branch. Hot-flasher evolutionary tracks are shown in Figure$\,$\ref{fig7}, including three stellar surface mixing: He-flasher with no He enrichment, He-flasher with shallow mixing, and He-flasher with deep mixing.
The tracks match with the location of He-rich sdB stars better than He-sdO stars in our sample, which suggests that the hot-flasher scenario is more reasonable for He-sdB stars. Although they cannot explain the He-rich stars, because the results of \cite{Miller2008} indicate that the lifetime from the core He flash to the ZAHB is around $2\times10^6$ yr which is far shorter than that of the He-core burning stage ($65-90\times10^6$ yr).

 \subsection{Effective temperature and helium abundance}
\begin{figure*}
\epsscale{0.9}
\plotone{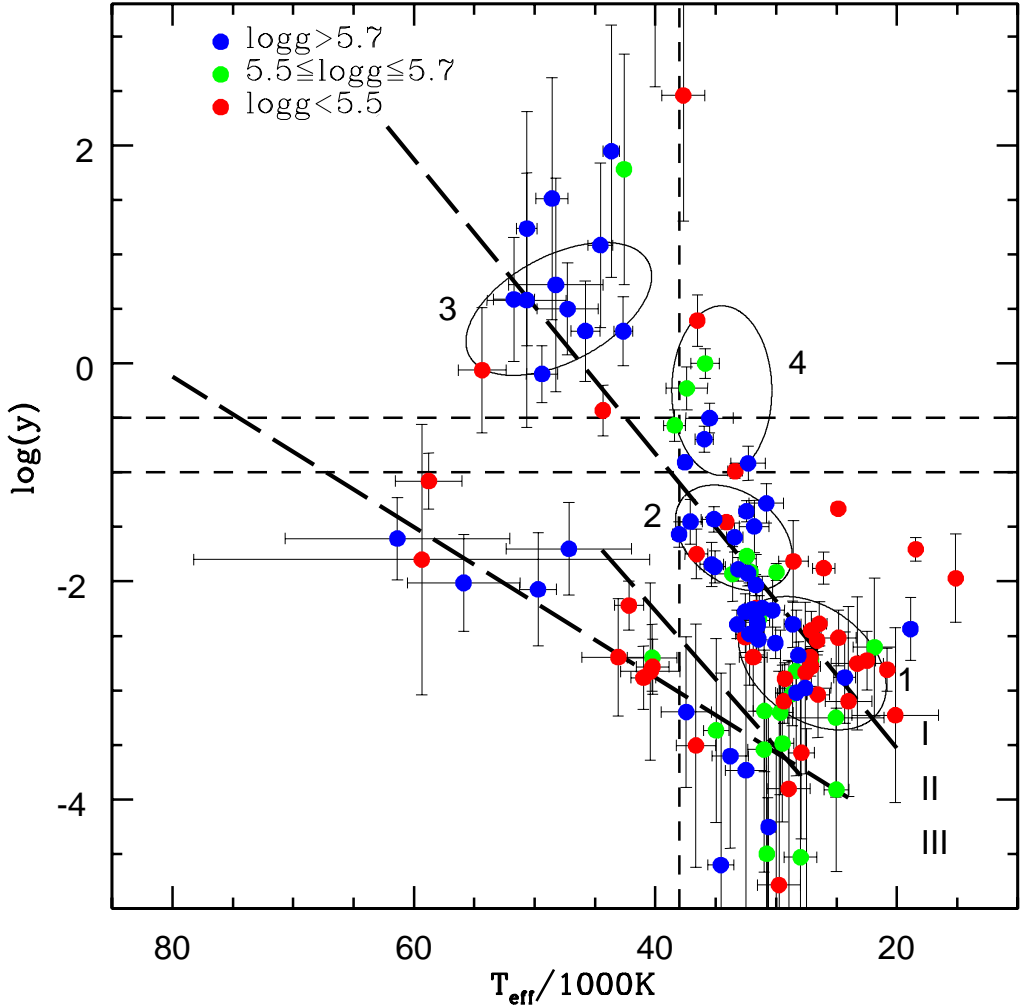}
\caption{Helium abundance versus effective temperature.
The long dashed lines are the best fit of the two sdB sequences from \cite{Edelmann2003} and one sdO sequence from \cite{Nemeth2012}. Four thin dashed lines denote $\log(y)=-0.5$, $\log(y)=-1$, $\log(y)=-4$ and $T_{\rm{eff}}=38000\rm{K}$. Ellipses 1, 2 and 3 are similar to those in \cite{Nemeth2012} and ellipse 4 shows the clustering of the He-rich sdB stars.\label{fig8}}
\end{figure*}

The He abundance plays a key role in understanding the formation and evolution of hot subdwarf stars. The effective temperature and He abundance plane is another important parameter space in looking for the evolutionary links between sdB and sdO stars.
The distribution of stars in the $T_{\rm{eff}}-\log(y)$ panel is shown in Figure$\,$\ref{fig8}.
We can see that sdB and sdO stars form two sequences having clear trends: at higher temperatures they have higher He abundances on average. In order to compare with previous results, we plot the two best-fit trends for sdB stars from \cite{Edelmann2003}:
\begin{equation}
\rm{I:}
\log(y)=-3.53+1.35(T_{\rm{eff}}/10^{4}\rm{K}-2.00),
\end{equation}
\begin{equation}
\rm{II:}
\log(y)=-4.79+1.26(T_{\rm{eff}}/10^{4}\rm{K}-2.00)
\end{equation}
and the one for sdO stars from \cite{Nemeth2012}:
\begin{equation}
\rm{III:}
 \log(y)=-4.26+0.69(T_{\rm{eff}}/10^{4}\rm{K}-2.00).
\end{equation}
First, the majority of our stars lie near or above the first sequence.
The first best-fitting trend is able to match the He-deficient sdB stars in the first sequence, but cannot fit the He-rich ones. This is in agreement with the report of \cite{Hirsch2009}. One can deduce that He-deficient and He-rich stars may origin from different formation channels.
Unlike sdB stars, He-rich sdO stars are more dispersive above the first sequence.
Although it seems that some He-rich sdO stars lie in the extension of the first best-fit line,
He-rich sdO stars with $\log{y}>0$ appear to follow a probable contrary tendency having higher temperatures and a lower helium abundance, which is consistent with the observations of \cite{Stroeer2007}.
In the second sequence, He-deficient sdO and sdB stars are found.
Except three sdB stars with $\log(y)<-4$, all stars in the second sequence can be matched with the third best-fitting trend.
In the range of the second trend line we have a few sdB stars, but they show too large scatter to be associated with the trend. As reported in the \cite{Nemeth2012}, there are different correlations for sdB and sdO stars, and the distribution of stars is more complex than linear trends. To date, its nature is still uncertain though it has appeared in many observations \citep{Edelmann2003, Lisker2005, OToole2008, Nemeth2012}.

Besides the above sequences, a clustering of He-rich sdB stars was found and marked in Figure$\,$\ref{fig8} by an ellipse (no.4). They are separated from sdB stars by a gap in the He abundance round $\log(y)=-1$ and from sdO stars by another gap in the temperature near $T_{\rm eff}=40\,000$K.
\cite{Nemeth2012} reported only five stars in this region from the GALEX sample. As the GALEX sample was limited to bright stars ($V<15$ mag) and our sample reach deeper the most likely reason for the clustering of He-rich sdB stars is that the LAMOST sample has a mixture of the thin-disk and thick-disk populations of hot subdwarfs.
As reported above, these stars do not follow the best-fit line of He-deficient sdB stars in the first sequence and the differences are very obvious. These also suggest that they probably origin from different formation channels and/or belong to different populations. The other three groups, similar to group 1, 2 and 3 in \cite{Nemeth2012} can also be seen in Figure$\,$\ref{fig8}.
As reported in \cite{Nemeth2012}, sdO stars show a gap between $\log(y)>-1.5$ and $\log(y)<-0.5$ and abundance extremes exist. They are either He-rich or He-deficient and we observe only a few stars in the $ -1 < \log(y) < 0$ abundance range. With increasing temperature these extremes become less prominent and the He abundance approaches to $\log(y)\sim-0.5$.
In addition, one can also see that most stars with $\log(g)<5.5$ crowd around $\log{y}=-2.7$, while other stars with $\log(g)\le5.5$ are scattered in the whole region, which is also similar to the results of \cite{Nemeth2012}.

\begin{figure*}
\epsscale{0.9}
\plotone{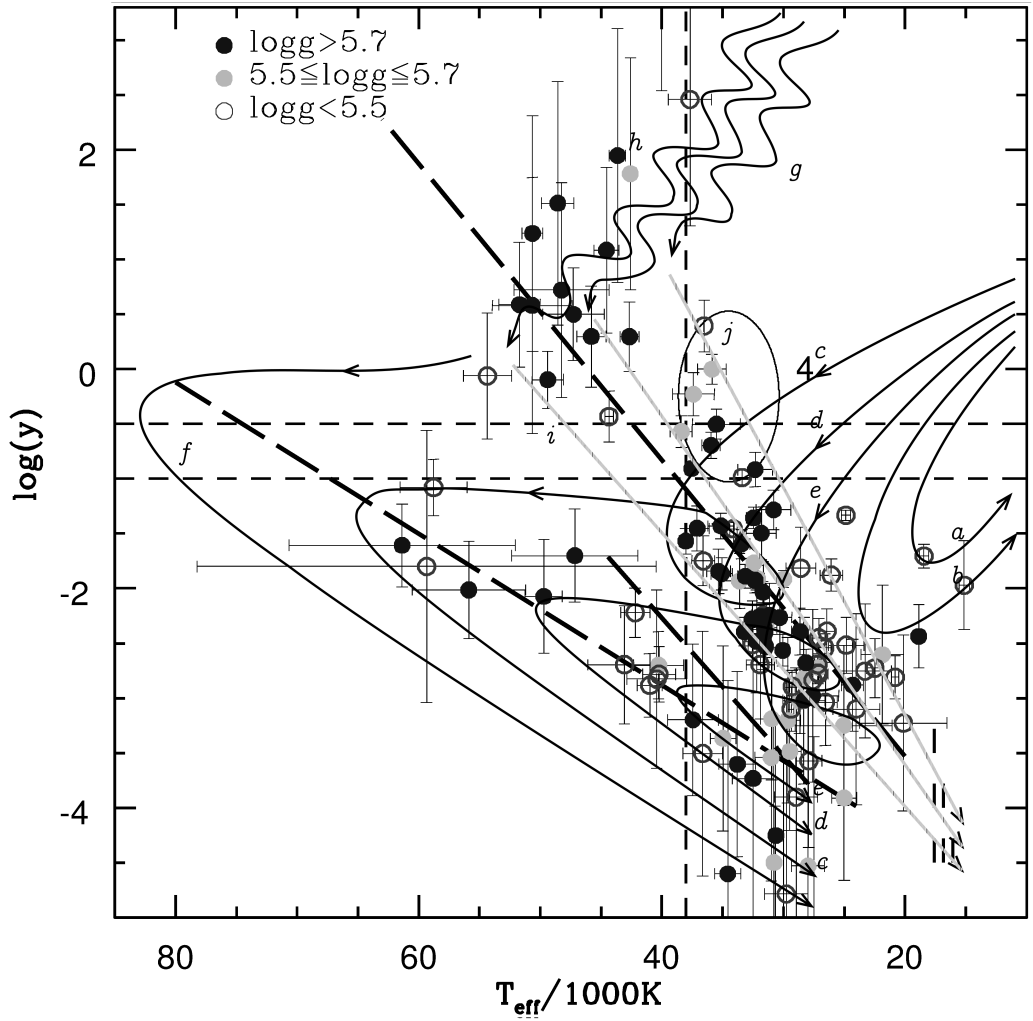}
\caption{Possible evolutionary sketches for hot subdwarf stars through the canonical evolution and double WDs merger channels \citep{Nemeth2012}.\label{fig9}}
\end{figure*}

There are no formation and evolution models for hot subdwarf stars that are able to make detailed calculations for the evolution of the surface He abundances and would allow for a direct comparison with observations in the $T_{\rm{eff}}-\log(y)$ panels.
Therefore, we make just a simple comparison with evolutionary sketches derived by \cite{Nemeth2012} based on observations and theoretical predictions.
Figure \ref{fig9} shows the canonical scenarios \citep{Han2002,Zhang2012} and Figure \ref{fig10} displays the hot-flasher scenarios \citep{Miller2008}. In Figure \ref{fig9}, lines a and b represent the canonical evolution of BHB stars. The He abundance decreases with surface temperature until core He burning is on. When core He burning stops, these stars evolve towards the AGB at lower surface temperatures and the He abundance gradually increases.
Lines c, d and e denote the canonical evolution of sdB stars. The He sinks to about $24000\,\rm{K}$. Over $24000\,\rm{K}$ the increasing UV flux starts a steady increase of the surface He abudance until about $36000-38000\,\rm{K}$ where core He burning exhausts \citep{OToole2008}. Next, a He shell-burning episode stars in the post-EHB phase. After passing the post-EHB, stars reach a maximum temperature and rapidly evolve to WDs and He sinks again. Lines g and h show the evolution of He-rich sdO stars via the slow and fast double WDs merger channels \citep{Zhang2012}. Lines i, j and f represent the evolution of He-rich sdO stars following the core He exhaustion. When core He burning stops, stars either evolve directly to lower temperatures and He abundances or reach a maximum temperature before evolving to WDs.
In Figure \ref{fig10}, lines k and m are from the predictions of the hot-flasher scenarios \citep{Miller2008}.
Line m shows a proposed evolutionary link between He-rich sdO and He-rich sdB stars.
Line k denotes larger loops from He shell flashes than the canonical theory
during the evolution of He-rich sdO stars. The canonical evolution (lines c, d and e) is also shown in Figure \ref{fig10} as the dominant channel to which the hot-flasher scenario contributes.
We can see that the canonical evolutionary tracks (lines c, d and e) can successfully describe the He-deficient sdB stars in the first sdB sequence and associates the formation of He-deficient sdO stars with evolved sdB stars as their successors.
The second sdB sequence may need to be reconsidered because it represents an intermediate evolutionary stage of sdB stars after a core He exhaustion. But it needs to be confirmed with a larger and more complete sample. The third sequence is a limit, no subdwarfs are observed at a higher temperature and gravity than this.
The double WDs merger channels (lines g and h) could explain the distribution of He-rich sdO stars and the evolution of surface He abundance from He-sdB to He-rich sdO stars. These are consistent with our results derived from the above $T_{\rm{eff}}-\log(g)$ plane. The hot-flasher scenario (lines \textit{k} and \textit{m}) could provide a possible explanation for He-rich sdO stars, but it fails to interpret the clustering of He-rich sdO stars in the $T_{\rm{eff}}-\log(g)$ planes.
The hot-flasher evolutionary tracks cover He-rich sdB stars not only in the $T_{\rm{eff}}-\log(g)$ planes but also in the $T_{\rm{eff}}-\log(y)$ panels.
However, they do not show a good interpretation for the clustering of He-rich sdB stars in $T_{\rm{eff}}-\log(y)$ panels.
Their atmospheric parameters are similar to those of blue hook (BHk) stars in globular clusters (GC) \citep{Moehler2004}.
A recent result by \cite{Lei2015} shows that tidally enhanced stellar wind in binary evolution is able to naturally provide the huge mass loss on the RGB needed for the hot flasher scenario and it is a possible and reasonable formation channel for BHk stars in GCs.
We conclude that the hot-flasher scenario can provide a plausible interpretation for
He-rich sdB stars and explain some loops in the region of He-rich sdB stars.
Identifying such stars in composite spectra binaries and deriving precise abundance patterns would help understanding these stars.

Turbulent atmospheric mixing makes the tracking of formation theories difficult and might be responsible for the atmospheric properities of He-sdO stars \citep{OToole2008, Nemeth2012}. Figure \ref{fig11} shows the other possible evolutionary sketches based on their observations and theoretical predictions related to atmospheric mixing and stellar winds. The detailed explanation of the models can be found in \cite{Nemeth2012}.

The reasons  for the correlation of the helium abundance with temperature and the different structure of the sequences in the $T_{\rm eff}-\log(y)$ panel are not fully understood. Although, the evolutionary sketches of \cite{Nemeth2012} are able to provide a qualitative picture, which should be explored with numerical models. Based on observations,
\cite{OToole2008} presented two hypotheses, one explaining the trend in helium abundance with effective temperature using the known physics, and the other suggesting that the two separate trends are from two different yet related populations (post-RGB evolution). These hypotheses have far reaching implications for our understanding of hot subdwarf evolution. Most recently, \cite{Geier2013a} made a detailed discussion and suggested that the close binary hypothesis \citep{Aznar2002} cannot explain the helium sequences and the post-RGB evolution  \citep{OToole2008} is not able to explain all of the observations in a consistent way. Therefore, further spectroscopic observations will be needed to help resolving these problems.'

\subsection{Luminosity and helium abundance}
The luminosity distribution function (LDF) is an important tool in comparing the predictions of theoretical models to observations \citep{Lisker2005}. Although our sample suffers form some selection effects, and therefore we cannot directly compare it with the predictions of theoretical models, some important properties can be obtained from the $\log(L/L_{\rm{edd}})-\log(y)$ plane which we show in Fig.\ref{fig12}. The locations of ZAEHB, TAEHB and terminal age post-EHB (TAPEHB) \citep{Nemeth2012} are marked in the figure.

From the $\log(L/L_{\rm{edd}})-\log(y)$ plane we can see that most sdB and sdO stars lie in a narrow strip where the He abundance increases with the average luminosity. This suggests that both sdB and sdO stars may follow a correlation in the $\log(L/L_{\rm{edd}})-\log(y)$ plane. There is a possible sequence that not only continuously connects He-sdB, He-rich sdO and He-rich sdB stars but also extends to He-deficient sdB stars, which suggests that there is an evolutionary link among them as the predictions of the hot-flasher channels \citep{Miller2008} in Figure \ref{fig10}.
Moreover, most He-rich sdO stars crowd around in a region between TAEHB and TAPEHB but the other three stars are scattered in a wider region on the right of TAPEHB and are possible post-EHB stars. Whereas, He-deficient sdO stars concentrate near the TAPEHB and look like the continuous extension of He-deficient sdB stars in luminosity, which is consistent with the prediction of the canonical EHB models \citep{Dorman1993} that
He-deficient sdB stars are evolving towards He-deficient sdO stars.

In addition, there is one He-sdB stars near the TAPEHB that is similar to He-rich sdO stars in the $\log(L/L_{\rm{edd}})-\log(y)$ plane. This suggests that there is probably an evolutionary link between He-sdB and He-rich sdO stars, which is also in agreement with the predictions of the double WDs merger channels \citep{Zhang2012} that He-sdB stars are evolving towards He-rich sdO stars.

\section{Conclusions}
\begin{figure*}
\epsscale{0.9}
\plotone{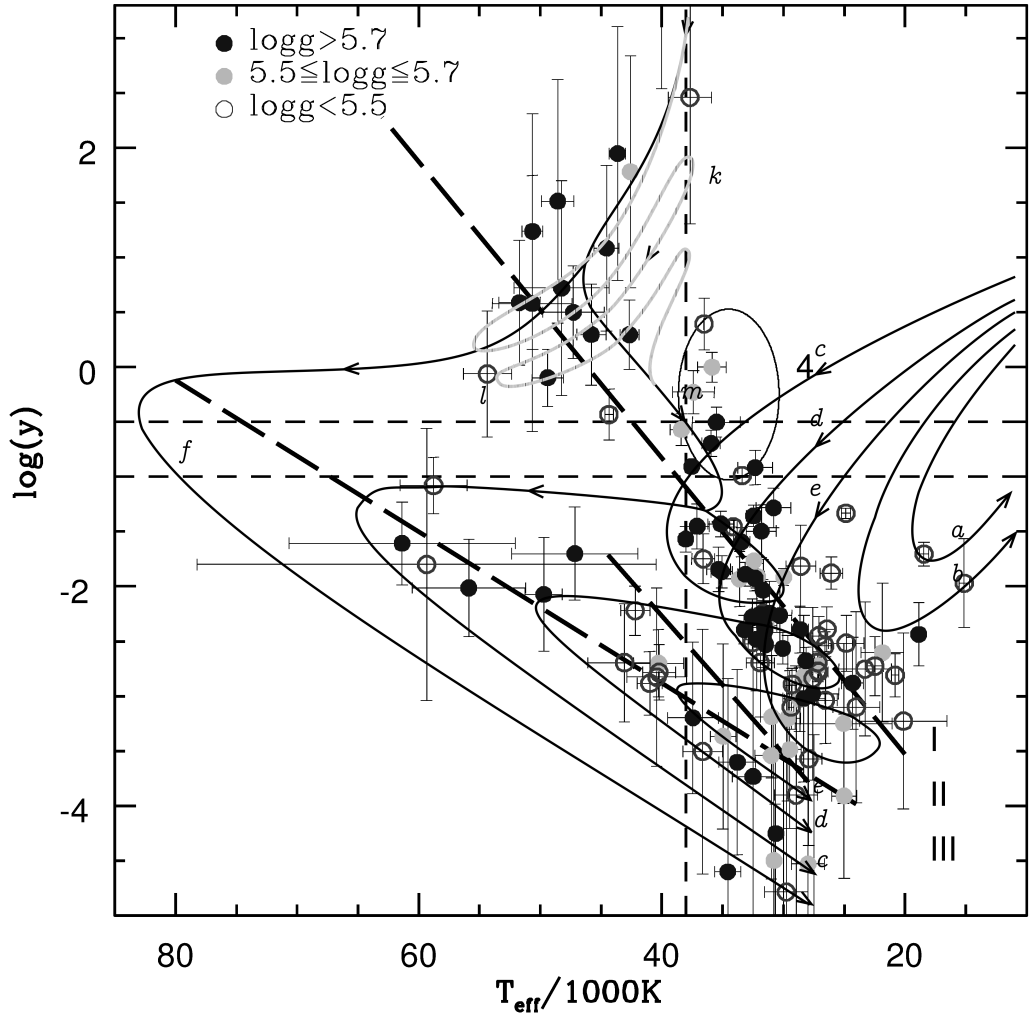}
\caption{Possible evolutionary sketches for hot subdwarf stars through the hot-flasher scenario \citep{Nemeth2012}.
\label{fig10}}
\end{figure*}

\begin{figure*}
\epsscale{0.9}
\plotone{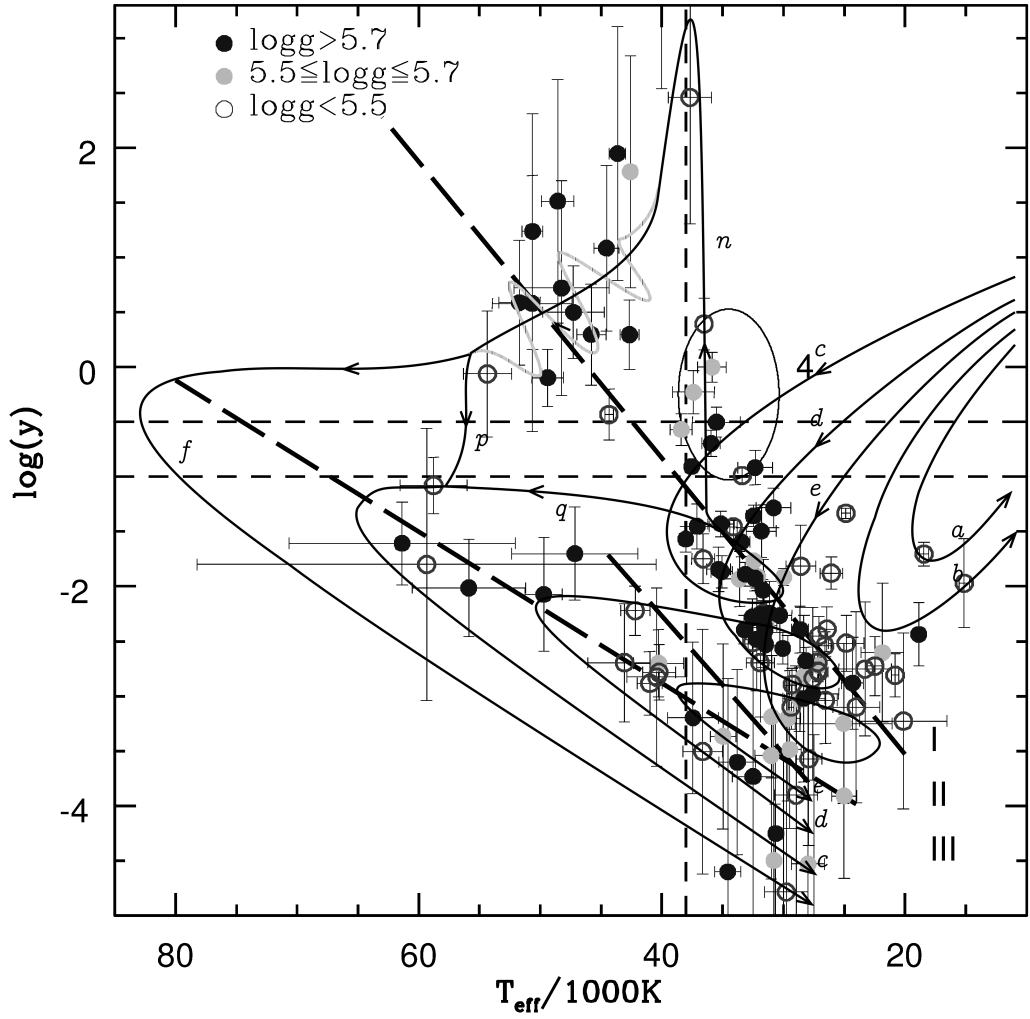}
\caption{Possible evolutionary sketches for hot subdwarf stars with atmospheric mixing and stellar winds  \citep{Nemeth2012}.
\label{fig11}}
\end{figure*}

\begin{figure*}
\epsscale{0.9}
\plotone{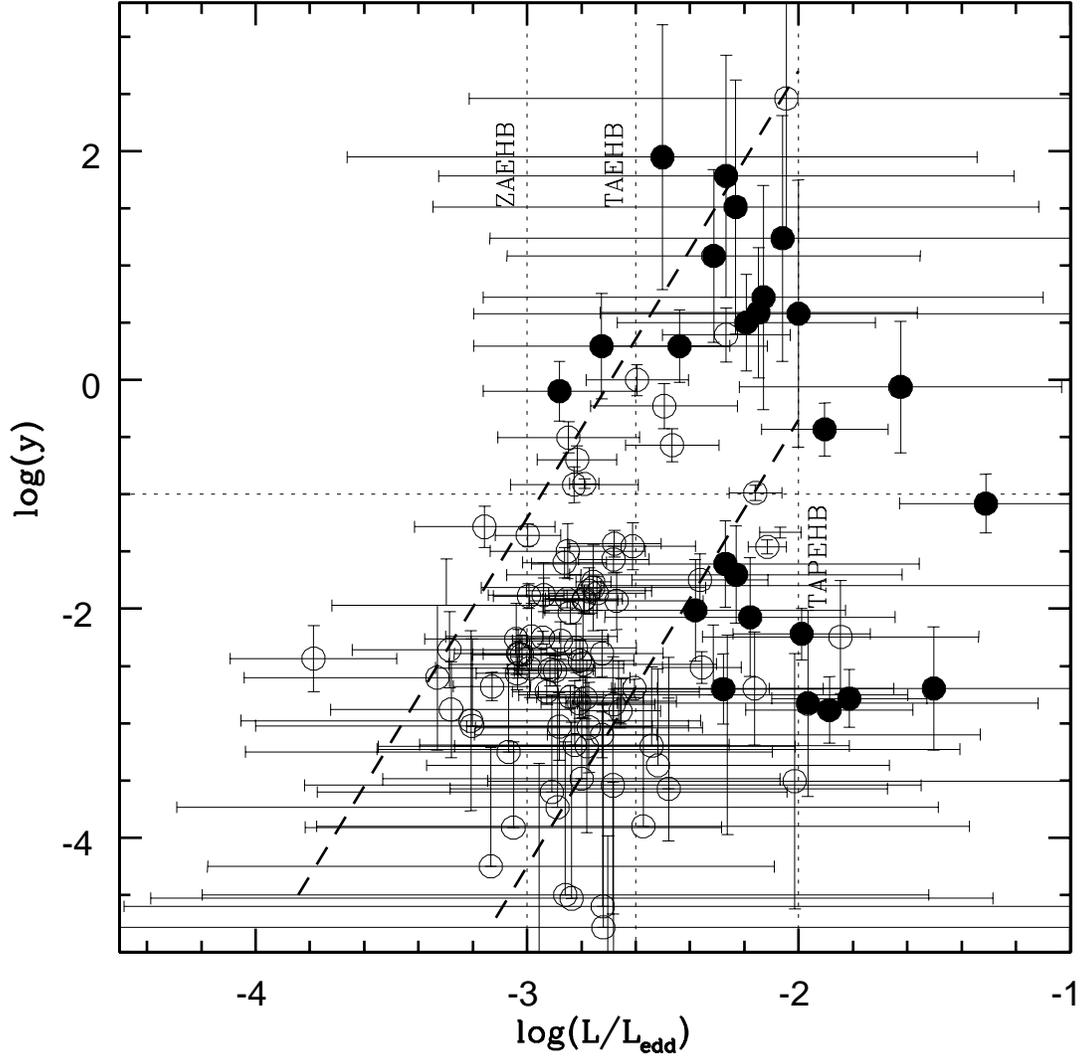}
\caption{Helium abundance versus luminosity with respect to the Eddington luminosity. The open circles represent sdB stars and the filled circles denote sdO stars. The location of the ZAEHB, TAEHB and terminal-age post-EHB (TAPEHB) \citep{Nemeth2012} are marked by the dotted lines. \label{fig12}}
\end{figure*}

We have identified 166 hot subdwarf stars from the spectra of LAMOST DR1 by using SDSS colours and catalogs of archive hot subdwarf stars, among which 44 stars show spectral signatures of cool companions in the observed optical spectra.
We have measured the atmospheric parameters (effective temperature $T_{\rm{eff}}$, surface gravity $\log(g)$, and He abundances $y=n(\rm{He})/n(\rm{H})$) of 122 non-composite stars by simultaneously fitting the profiles of H and He lines using synthetic spectra calculated from non-LTE {\sc Tlusty} model atmospheres. 27 stars are classified as sdO stars and 88 as sdB stars.
The LAMOST sample properties have been obtained and compared to various formation channels in the $T_{\rm eff} - \log(g)$, $T_{\rm eff} - log(y)$ and $\log(L/L_{\rm edd}) - log(y)$ planes. The evolutionary status of the stars has been discussed based on the observations and theoretical predictions. The following conclusions can be drawn:
\begin{enumerate}

\item
In the $T_{\rm eff}-\log{g}$ plane, most of the sdB stars lie in the EHB band and two well defined groups can be outlined.
Binary population synthesis models predict distinct properties of subdwarfs from the different channels. Therefore it is tempting to associate the lower temperature and surface gravity group 1 with the Common-Envelope formation channel, and the higher temperature and gravity group 2 with the Roche-lobe Overflow channel. \cite{Nemeth2012} found that long-period composite spectra binaries (sdB+F/G) from the Roche-lobe overflow channel show up exclusively in the higher temperature and gravity group. However, observations \citep{Kawka2015, Kupfer2015} show that both short- and long-period binaries occur in each group, suggesting that they have a mixture of stars with different formation history. The existence of the two sdB groups in the temperature gravity plane is an important result, but further investigations are needed to find their significance and weather we can infer from these groups to the yield of various formation channels \citep{Han2003}. Therefore, we will revisit the significance of the sdB groups in a forthcoming publication where we also include the results from composite spectra binaries from LAMOST DR1. The sdO stars also show two groups, one is the known He-rich sdO stars near $T_{\rm{eff}}=45\,000\,\rm{K}$ and $\log(g)=5.8$, another is the mixture of the He-rich and He-deficient sdO and sdB stars around $T_{\rm{eff}}=38\,000\,\rm{K}$ and $\log(g)=5.3$. Both the canonical post-EHB evolution and the non-canonical hot-flasher scenario fail to explain the clustering of He-rich sdO stars, but in the canonical double WD binary merger channel is viable. Whereas, the non-canonical hot-flasher scenario can provide a reasonable option for the He-rich sdB stars.
If the evolution of surface He abundances is not taken into account, our sample can be interpreted with the canonical scenario. Our sample also supports that He-rich sdO stars are more frequent than the He-deficient ones \citep{Stroeer2007}.

\item
In the $T_{\rm{eff}}-\log(y)$ plane, sdB and sdO stars show two sequences. A clustering of He-rich sdB stars is found, but published trends \citep{Edelmann2003} for the first sequence is not suitable for them, which is in agreement with results of \cite{Hirsch2009}.
In addition, this group of stars is missing from the sample of \cite{Nemeth2012}.
Moreover, we see that sdO stars display a big gap in the $ -1.5 < \log(y) < 0.5$ abundance range.
The sdO stars also show abundance extremes, they are either He-rich or He-deficient and we observe only a few stars in the $ -1 < \log(y) < 0$ abundance range. With increasing temperature these extremes become less prominent and the He abundance approaches to $\log(y)\sim-0.5$.
We have compared our results with evolutionary sketches derived by \cite{Nemeth2012} by comparing their observations to theoretical predictions. The evolutionary sketches for the canonical scenarios can cover all stars except He-rich sdB stars.
These suggests that He-rich and He-deficient sdB stars may origin from different formation channels.
We deduced that the second sequence represents the post-EHB stage.
Although the evolutionary sketches for the hot-flasher scenario, as a non-canonical scenario, can also cover He-rich sdB stars, they may not explain the clustering of He-rich sdB stars. A recent work \cite{Lei2015} shows that the hot-flasher scenario seems to provide a plausible interpretation for them in cases where tidally enhanced stellar wind in binary evolution is taken into account.
We can also deduce that there are some evolutionary loops in the region of He-rich sdB stars that can be associated with the hot-flasher scenario.

\item
In the $\log(L/L_{\rm edd})-\log(y)$, one can find that most sdB and sdO stars lie in a narrow strip. This indicates that there may be follow a correlation between sdB and sdO stars. They show a possible sequence that not only continuously connects He-rich sdO and He-rich sdB stars but also extends to He-deficient sdB stars, which suggests that there is an evolutionary link among them as predicted by the hot-flasher channels \citep{Miller2008} in Figure \ref{fig10}. We also find that He-deficient sdO stars concentrate near the TAPEHB and look like the continuous extension of He-deficient sdB stars in luminosity, which implies that sdB stars evolve into He-deficient sdO stars. Whereas, He-rich sdO stars are found in a wider luminosity region. Most of them crowd around a region between the TAEHB and TAPEHB but the other three stars are scattered in a wider region on the right of the TAPEHB and are possible post-EHB stars.
There is one He-sdB star that is similar to He-rich sdO stars in $\log(L/L_{\rm{edd}})-\log(y)$ panel, which suggests that there is an evolutionary link among them as predicted by the double WD binary merger channels.
These results are consistent with previous studies \citep{Edelmann2003, Lisker2005, Stroeer2007, Hirsch2009, Nemeth2012, Geier2013, Geier2013a,Geier2015}.
\end{enumerate}

\acknowledgments
P.N. was supported by the Deutsche Forschungsgemeinschaft under grant He 1356/49-2.
We acknowledge the Strategic Priority Research Program The Emergence of Cosmological Structures of the Chinese Academy of Sciences, Grant No. XDB09000000, the National Key Basic Research Program of China 2014CB845700, the National Natural Science Foundation of China (NSFC) grants 11303021, 11390374, 11373032, 11333003, 11473037, 11521303£¬ U1231202, and U1231119, and the Foundation of China West Normal University, Grant No. 12B206 and CXTD2014-1.
Guoshoujing Telescope (the Large Sky Area Multi-Object Fiber Spectroscopic Telescope LAMOST) is a National Major Scientific Project built by the Chinese Academy of Sciences. Funding for the project has been provided by the National Development and Reform Commission. LAMOST is operated and managed by the National Astronomical Observatories, Chinese Academy of Sciences.
This research has made use of the VizieR catalogue access tool, CDS, Strasbourg, France.


\begin{thebibliography}{65}
\expandafter\ifx\csname natexlab\endcsname\relax\def\natexlab#1{#1}\fi

\bibitem[{{Aznar Cuadrado} \& {Jeffery}(2002)}]{Aznar2002}
{Aznar Cuadrado}, R., \& {Jeffery}, C.~S. 2002, \aap, 385, 131

\bibitem[{{Carlin} {et~al.}(2012){Carlin}, {L{\'e}pine}, {Newberg}, {Deng},
  {Beers}, {Chen}, {Christlieb}, {Fu}, {Gao}, {Grillmair}, {Guhathakurta},
  {Han}, {Hou}, {Lee}, {Li}, {Liu}, {Liu}, {Pan}, {Sellwood}, {Wang}, {Yang},
  {Yanny}, {Zhang}, {Zheng}, \& {Zhu}}]{Carlin2012}
{Carlin}, J.~L., {et~al.} 2012, Res. Astron. Astrophys., 12, 755

\bibitem[{{Charpinet} {et~al.}(2010){Charpinet}, {Green}, {Baglin}, {Van
  Grootel}, {Fontaine}, {Vauclair}, {Chaintreuil}, {Weiss}, {Michel},
  {Auvergne}, {Catala}, {Samadi}, \& {Baudin}}]{Charpinet2010}
{Charpinet}, S., {et~al.} 2010, \aap, 516, L6

\bibitem[{{Chen} {et~al.}(2012){Chen}, {Hou}, {Yu}, {Liu}, {Deng}, {Newberg},
  {Carlin}, {Yang}, {Zhang}, {Shen}, {Zhang}, {Chen}, {Chen}, {Christlieb},
  {Han}, {Lee}, {Liu}, {Pan}, {Shi}, {Wang}, \& {Zhu}}]{Chenl2012}
{Chen}, L., {et~al.} 2012, Res. Astron. Astrophys., 12, 805

\bibitem[{{Cui} {et~al.}(2012){Cui}, {Zhao}, {Chu}, {Li}, {Li}, {Zhang}, {Su},
  {Yao}, {Wang}, {Xing}, {Li}, {Zhu}, {Wang}, {Gu}, {Luo}, {Xu}, {Zhang},
  {Liu}, {Zhang}, {Yang}, {Cao}, {Chen}, {Chen}, {Chen}, {Chen}, {Chu}, {Feng},
  {Gong}, {Hou}, {Hu}, {Hu}, {Hu}, {Jia}, {Jiang}, {Jiang}, {Jiang}, {Jin},
  {Li}, {Li}, {Li}, {Liu}, {Liu}, {Lu}, {Mao}, {Men}, {Qi}, {Qi}, {Shi},
  {Tang}, {Tao}, {Wang}, {Wang}, {Wang}, {Wang}, {Wang}, {Wang}, {Wang},
  {Wang}, {Wang}, {Wang}, {Wang}, {Wang}, {Xu}, {Xu}, {Yang}, {Yu}, {Yuan},
  {Yuan}, {Zhai}, {Zhang}, {Zhang}, {Zhang}, {Zhao}, {Zhou}, {Zhou}, {Zhu}, \&
  {Zou}}]{Cui2012}
{Cui}, X.-Q., {et~al.} 2012, Res. Astron. Astrophys., 12, 1197

\bibitem[{{Cutri} {et~al.}(2003){Cutri}, {Skrutskie}, {van Dyk}, {Beichman},
  {Carpenter}, {Chester}, {Cambresy}, {Evans}, {Fowler}, {Gizis}, {Howard},
  {Huchra}, {Jarrett}, {Kopan}, {Kirkpatrick}, {Light}, {Marsh}, {McCallon},
  {Schneider}, {Stiening}, {Sykes}, {Weinberg}, {Wheaton}, {Wheelock}, \&
  {Zacarias}}]{Cutri2003}
{Cutri}, R.~M., {et~al.} 2003, VizieR Online Data Catalog, 2246, 0

\bibitem[{{D'Cruz} {et~al.}(1996){D'Cruz}, {Dorman}, {Rood}, \&
  {O'Connell}}]{DCruz1996}
{D'Cruz}, N.~L., {Dorman}, B., {Rood}, R.~T., \& {O'Connell}, R.~W. 1996, \apj,
  466, 359

\bibitem[{{Deng} {et~al.}(2012){Deng}, {Newberg}, {Liu}, {Carlin}, {Beers},
  {Chen}, {Chen}, {Christlieb}, {Grillmair}, {Guhathakurta}, {Han}, {Hou},
  {Lee}, {L{\'e}pine}, {Li}, {Liu}, {Pan}, {Sellwood}, {Wang}, {Wang}, {Yang},
  {Yanny}, {Zhang}, {Zhang}, {Zheng}, \& {Zhu}}]{Deng2012}
{Deng}, L.-C., {et~al.} 2012, Res. Astron. Astrophys., 12, 735

\bibitem[{{Dorman} {et~al.}(1993){Dorman}, {Rood}, \& {O'Connell}}]{Dorman1993}
{Dorman}, B., {Rood}, R.~T., \& {O'Connell}, R.~W. 1993, \apj, 419, 596

\bibitem[{{Drilling} {et~al.}(2013){Drilling}, {Jeffery}, {Heber}, {Moehler},
  \& {Napiwotzki}}]{Drilling2013}
{Drilling}, J.~S., {Jeffery}, C.~S., {Heber}, U., {Moehler}, S., \&
  {Napiwotzki}, R. 2013, \aap, 551, A31

\bibitem[{{Drilling} {et~al.}(2003){Drilling}, {Moehler}, {Jeffery}, {Heber},
  \& {Napiwotzki}}]{Drilling2003}
{Drilling}, J.~S., {Moehler}, S., {Jeffery}, C.~S., {Heber}, U., \&
  {Napiwotzki}, R. 2003, in The Garrison Festschrift, ed. R.~O. {Gray}, C.~J.
  {Corbally}, \& A.~G.~D. {Philip}, 27

\bibitem[{{Edelmann} {et~al.}(2003){Edelmann}, {Heber}, {Hagen}, {Lemke},
  {Dreizler}, {Napiwotzki}, \& {Engels}}]{Edelmann2003}
{Edelmann}, H., {Heber}, U., {Hagen}, H.-J., {Lemke}, M., {Dreizler}, S.,
  {Napiwotzki}, R., \& {Engels}, D. 2003, \aap, 400, 939

\bibitem[{{Fontaine} {et~al.}(2012){Fontaine}, {Brassard}, {Charpinet},
  {Green}, {Randall}, \& {Van Grootel}}]{Fontaine2012}
{Fontaine}, G., {Brassard}, P., {Charpinet}, S., {Green}, E.~M., {Randall},
  S.~K., \& {Van Grootel}, V. 2012, \aap, 539, A12

\bibitem[{{Geier}(2013)}]{Geier2013}
{Geier}, S. 2013, \aap, 549, A110

\bibitem[{{Geier} {et~al.}(2013){Geier}, {Heber}, {Edelmann}, {Morales-Rueda},
  {Kilkenny}, {O'Donoghue}, {Marsh}, \& {Copperwheat}}]{Geier2013a}
{Geier}, S., {Heber}, U., {Edelmann}, H., {Morales-Rueda}, L., {Kilkenny}, D.,
  {O'Donoghue}, D., {Marsh}, T.~R., \& {Copperwheat}, C. 2013, \aap, 557, A122

\bibitem[{{Geier} {et~al.}(2007){Geier}, {Nesslinger}, {Heber}, {Przybilla},
  {Napiwotzki}, \& {Kudritzki}}]{Geier2007}
{Geier}, S., {Nesslinger}, S., {Heber}, U., {Przybilla}, N., {Napiwotzki}, R.,
  \& {Kudritzki}, R.-P. 2007, \aap, 464, 299

\bibitem[{{Geier} {et~al.}(2011){Geier}, {Hirsch}, {Tillich}, {Maxted},
  {Bentley}, {{\O}stensen}, {Heber}, {G{\"a}nsicke}, {Marsh}, {Napiwotzki},
  {Barlow}, \& {O'Toole}}]{Geier2011}
{Geier}, S., {et~al.} 2011, \aap, 530, A28

\bibitem[{{Geier} {et~al.}(2015){Geier}, {Kupfer}, {Heber}, {Schaffenroth},
  {Barlow}, {{\O}stensen}, {O'Toole}, {Ziegerer}, {Heuser}, {Maxted},
  {G{\"a}nsicke}, {Marsh}, {Napiwotzki}, {Br{\"u}nner}, {Schindewolf}, \&
  {Niederhofer}}]{Geier2015}
---. 2015, \aap, 577, A26

\bibitem[{{Green} {et~al.}(1986{\natexlab{a}}){Green}, {Schmidt}, \&
  {Liebert}}]{Green1986}
{Green}, R.~F., {Schmidt}, M., \& {Liebert}, J. 1986{\natexlab{a}}, \apjs, 61,
  305

\bibitem[{{Green} {et~al.}(1986{\natexlab{b}}){Green}, {Schmidt}, \&
  {Liebert}}]{pg1986}
---. 1986{\natexlab{b}}, \apjs, 61, 305

\bibitem[{{Han}(2008)}]{Han2008}
{Han}, Z. 2008, \aap, 484, L31

\bibitem[{{Han} {et~al.}(2010){Han}, {Jeffery}, {Podsiadlowski}, \&
  {Dopita}}]{Han2010}
{Han}, Z., {Jeffery}, S., {Podsiadlowski}, P., \& {Dopita}, M.~A. 2010, \apss,
  329, 1

\bibitem[{{Han} {et~al.}(1994){Han}, {Podsiadlowski}, \& {Eggleton}}]{Han1994}
{Han}, Z., {Podsiadlowski}, P., \& {Eggleton}, P.~P. 1994, \mnras, 270, 121

\bibitem[{{Han} {et~al.}(2007){Han}, {Podsiadlowski}, \&
  {Lynas-Gray}}]{Han2007}
{Han}, Z., {Podsiadlowski}, P., \& {Lynas-Gray}, A.~E. 2007, \mnras, 380, 1098

\bibitem[{{Han} {et~al.}(2003){Han}, {Podsiadlowski}, {Maxted}, \&
  {Marsh}}]{Han2003}
{Han}, Z., {Podsiadlowski}, P., {Maxted}, P.~F.~L., \& {Marsh}, T.~R. 2003,
  \mnras, 341, 669

\bibitem[{{Han} {et~al.}(2002){Han}, {Podsiadlowski}, {Maxted}, {Marsh}, \&
  {Ivanova}}]{Han2002}
{Han}, Z., {Podsiadlowski}, P., {Maxted}, P.~F.~L., {Marsh}, T.~R., \&
  {Ivanova}, N. 2002, \mnras, 336, 449

\bibitem[{{Heber}(2009)}]{Heber2009}
{Heber}, U. 2009, \araa, 47, 211

\bibitem[{{Heber} {et~al.}(1984){Heber}, {Hunger}, {Jonas}, \&
  {Kudritzki}}]{Heber1984}
{Heber}, U., {Hunger}, K., {Jonas}, G., \& {Kudritzki}, R.~P. 1984, \aap, 130,
  119

\bibitem[{{Hirsch}(2009)}]{Hirsch2009}
{Hirsch}, H. 2009, PhD thesis, University of Erlangen-N$\ddot{u}$rnberg

\bibitem[{{Hubeny} \& {Lanz}(1995)}]{hubeny95}
{Hubeny}, I., \& {Lanz}, T. 1995, \apj, 439, 875

\bibitem[{{Jester} {et~al.}(2005){Jester}, {Schneider}, {Richards}, {Green},
  {Schmidt}, {Hall}, {Strauss}, {Vanden Berk}, {Stoughton}, {Gunn},
  {Brinkmann}, {Kent}, {Smith}, {Tucker}, \& {Yanny}}]{Jester2005}
{Jester}, S., {et~al.} 2005, \aj, 130, 873

\bibitem[{{Justham} {et~al.}(2011){Justham}, {Podsiadlowski}, \&
  {Han}}]{Justham2011}
{Justham}, S., {Podsiadlowski}, P., \& {Han}, Z. 2011, \mnras, 410, 984

\bibitem[{{Kupfer}  {et~al.}(2015){Kupfer}, {Geier}, \&
 {Heber}}]{Kupfer2015}
 {Kupfer}, T., {Geier}, S., {Heber}, U., {et~al.} 2015, \aap, 576, A44

\bibitem[{{Kawka} {et~al.}(2015){Kawka}, {Vennes}, \&
{O'Toole}}]{Kawka2015}
{Kawka}, A., {Vennes}, S., {O'Toole}, S., {et~al.}  2015, \mnras, 450, 3514

\bibitem[{{Lanz} {et~al.}(2004){Lanz}, {Brown}, {Sweigart}, {Hubeny}, \&
  {Landsman}}]{Lanz2004}
{Lanz}, T., {Brown}, T.~M., {Sweigart}, A.~V., {Hubeny}, I., \& {Landsman},
  W.~B. 2004, \apj, 602, 342

\bibitem[{{Lanz} \& {Hubeny}(2007)}]{lanz07}
{Lanz}, T., \& {Hubeny}, I. 2007, \apjs, 169, 83

\bibitem[{{Lasker} {et~al.}(2008){Lasker}, {Lattanzi}, {McLean}, {Bucciarelli},
  {Drimmel}, {Garcia}, {Greene}, {Guglielmetti}, {Hanley}, {Hawkins},
  {Laidler}, {Loomis}, {Meakes}, {Mignani}, {Morbidelli}, {Morrison},
  {Pannunzio}, {Rosenberg}, {Sarasso}, {Smart}, {Spagna}, {Sturch},
  {Volpicelli}, {White}, {Wolfe}, \& {Zacchei}}]{Lasker2008}
{Lasker}, B.~M., {et~al.} 2008, \aj, 136, 735

\bibitem[{{Lei} {et~al.}(2015){Lei}, {Chen}, {Zhang}, \& {Han}}]{Lei2015}
{Lei}, Z., {Chen}, X., {Zhang}, F., \& {Han}, Z. 2015, \mnras, 449, 2741

\bibitem[{{Lei} {et~al.}(2013){Lei}, {Chen}, {Zhang}, \& {Han}}]{Lei2013}
{Lei}, Z.-X., {Chen}, X.-F., {Zhang}, F.-H., \& {Han}, Z. 2013, \aap, 549, A145

\bibitem[{{Lemke}(1997)}]{lemke97}
{Lemke}, M. 1997, \aaps, 122, 285

\bibitem[{{Lisker} {et~al.}(2005){Lisker}, {Heber}, {Napiwotzki}, {Christlieb},
  {Han}, {Homeier}, \& {Reimers}}]{Lisker2005}
{Lisker}, T., {Heber}, U., {Napiwotzki}, R., {Christlieb}, N., {Han}, Z.,
  {Homeier}, D., \& {Reimers}, D. 2005, \aap, 430, 223

\bibitem[{{Liu} {et~al.}(2014){Liu}, {Deng}, {Carlin}, {Smith}, {Li},
  {Newberg}, {Gao}, {Yang}, {Xue}, {Xu}, {Zhang}, {Xin}, {Wu}, \&
  {Jin}}]{Liu2014}
{Liu}, C., {et~al.} 2014, \apj, 790, 110

\bibitem[{{Luo} {et~al.}(2014){Luo}, {Zhang}, {Chen}, {Song}, {Wu}, {Bai},
  {Wang}, {Du}, \& {Zhang}}]{Luo2014b}
{Luo}, A., {et~al.} 2014, in IAU Symposium, Vol. 298, IAU Symposium, ed.
  S.~{Feltzing}, G.~{Zhao}, N.~A. {Walton}, \& P.~{Whitelock}, 428--428

\bibitem[{{Luo} {et~al.}(2012{\natexlab{a}}){Luo}, {Zhang}, {Zhao}, {Zhao},
  {Cui}, {Li}, {Chu}, {Shi}, {Wang}, {Zhang}, {Bai}, {Chen}, {Wang}, {Guo},
  {Chen}, {Du}, {Kong}, {Lei}, {Li}, {Song}, {Wu}, {Zhang}, {Zhou}, {Zuo},
  {Du}, {He}, {Hou}, {Dong}, {Li}, {Li}, {Li}, {Song}, {Tian}, {Wang}, {Wu},
  {Yang}, {Yuan}, {Cao}, {Chen}, {Chen}, {Chen}, {Chu}, {Feng}, {Gong}, {Gu},
  {Hou}, {Huo}, {Hu}, {Hu}, {Hu}, {Jia}, {Jiang}, {Jiang}, {Jiang}, {Jin},
  {Li}, {Li}, {Li}, {Li}, {Li}, {Liu}, {Liu}, {Liu}, {Lu}, {Lu}, {Luo}, {Mao},
  {Men}, {Ni}, {Qi}, {Qi}, {Shi}, {Su}, {Sun}, {Su}, {Tang}, {Tao}, {Tu},
  {Wang}, {Wang}, {Wang}, {Wang}, {Wang}, {Wang}, {Wang}, {Wang}, {Wang},
  {Wang}, {Wang}, {Wang}, {Wang}, {Wang}, {Wei}, {Xue}, {Xing}, {Xu}, {Xu},
  {Xu}, {Yang}, {Yang}, {Yao}, {Yu}, {Yuan}, {Zhai}, {Zhang}, {Zhang}, {Zhang},
  {Zhang}, {Zhang}, {Zhang}, {Zhao}, {Zhou}, {Zhu}, {Zhu}, \& {Zou}}]{Luo2012c}
{Luo}, A.-L., {et~al.} 2012{\natexlab{a}}, Res. Astron. Astrophys., 12, 1243

\bibitem[{{Luo} {et~al.}(2012{\natexlab{b}}){Luo}, {Zhang}, {Zhao}, {Zhao},
  {Cui}, {Li}, {Chu}, {Shi}, {Wang}, {Zhang}, {Bai}, {Chen}, {Wang}, {Guo},
  {Chen}, {Du}, {Kong}, {Lei}, {Li}, {Song}, {Wu}, {Zhang}, {Zhou}, {Zuo},
  {Du}, {He}, {Hou}, {Dong}, {Li}, {Li}, {Li}, {Song}, {Tian}, {Wang}, {Wu},
  {Yang}, {Yuan}, {Cao}, {Chen}, {Chen}, {Chen}, {Chu}, {Feng}, {Gong}, {Gu},
  {Hou}, {Huo}, {Hu}, {Hu}, {Hu}, {Jia}, {Jiang}, {Jiang}, {Jiang}, {Jin},
  {Li}, {Li}, {Li}, {Li}, {Li}, {Liu}, {Liu}, {Liu}, {Lu}, {Lu}, {Luo}, {Mao},
  {Men}, {Ni}, {Qi}, {Qi}, {Shi}, {Su}, {Sun}, {Su}, {Tang}, {Tao}, {Tu},
  {Wang}, {Wang}, {Wang}, {Wang}, {Wang}, {Wang}, {Wang}, {Wang}, {Wang},
  {Wang}, {Wang}, {Wang}, {Wang}, {Wang}, {Wei}, {Xue}, {Xing}, {Xu}, {Xu},
  {Xu}, {Yang}, {Yang}, {Yao}, {Yu}, {Yuan}, {Zhai}, {Zhang}, {Zhang}, {Zhang},
  {Zhang}, {Zhang}, {Zhang}, {Zhao}, {Zhou}, {Zhu}, {Zhu}, \& {Zou}}]{Luo2012}
---. 2012{\natexlab{b}}, Res. Astron. Astrophys., 12, 1243

\bibitem[{{Miller Bertolami} {et~al.}(2008){Miller Bertolami}, {Althaus},
  {Unglaub}, \& {Weiss}}]{Miller2008}
{Miller Bertolami}, M.~M., {Althaus}, L.~G., {Unglaub}, K., \& {Weiss}, A.
  2008, \aap, 491, 253

\bibitem[{{Moehler} {et~al.}(2004){Moehler}, {Sweigart}, {Landsman}, {Hammer},
  \& {Dreizler}}]{Moehler2004}
{Moehler}, S., {Sweigart}, A.~V., {Landsman}, W.~B., {Hammer}, N.~J., \&
  {Dreizler}, S. 2004, \aap, 415, 313

\bibitem[{{N{\'e}meth} {et~al.}(2012){N{\'e}meth}, {Kawka}, \&
  {Vennes}}]{Nemeth2012}
{N{\'e}meth}, P., {Kawka}, A., \& {Vennes}, S. 2012, \mnras, 427, 2180

\bibitem[{{Ochsenbein} {et~al.}(2000){Ochsenbein}, {Bauer}, \&
  {Marcout}}]{Ochsenbein2000}
{Ochsenbein}, F., {Bauer}, P., \& {Marcout}, J. 2000, \aaps, 143, 23

\bibitem[{{O'Connell}(1999)}]{Connell1999}
{O'Connell}, R.~W. 1999, \araa, 37, 603

\bibitem[{{{\O}stensen}(2004)}]{Ostensen2004}
{{\O}stensen}, R.~H. 2004, \apss, 291, 263

\bibitem[{{O'Toole}(2008)}]{OToole2008}
{O'Toole}, S.~J. 2008, in Astronomical Society of the Pacific Conference
  Series, Vol. 392, Hot Subdwarf Stars and Related Objects, ed. U.~{Heber},
  C.~S. {Jeffery}, \& R.~{Napiwotzki}, 67

\bibitem[{{Paczy{\'n}ski}(1971)}]{paczynski1971}
{Paczy{\'n}ski}, B. 1971, \actaa, 21, 1

\bibitem[{{Stroeer} {et~al.}(2007){Stroeer}, {Heber}, {Lisker}, {Napiwotzki},
  {Dreizler}, {Christlieb}, \& {Reimers}}]{Stroeer2007}
{Stroeer}, A., {Heber}, U., {Lisker}, T., {Napiwotzki}, R., {Dreizler}, S.,
  {Christlieb}, N., \& {Reimers}, D. 2007, \aap, 462, 269

\bibitem[{{Sweigart}(1997)}]{Sweigart1997}
{Sweigart}, A.~V. 1997, \apjl, 474, L23

\bibitem[{{Tremblay} \& {Bergeron}(2009)}]{tremblay11}
{Tremblay}, P.-E., \& {Bergeron}, P. 2009, \apj, 696, 1755

\bibitem[{{Wang} \& {Han}(2010)}]{Wang2010}
{Wang}, B., \& {Han}, Z. 2010, \aap, 515, A88

\bibitem[{{Webbink}(1984)}]{Webbink1984}
{Webbink}, R.~F. 1984, \apj, 277, 355

\bibitem[{{Winter}(2006)}]{Winter2006}
{Winter}, C. 2006, PhD thesis, Armagh Observatory

\bibitem[{{Xiang} {et~al.}(2015){Xiang}, {Liu}, {Yuan}, {Huang}, {Huo},
  {Zhang}, {Chen}, {Zhang}, {Sun}, {Wang}, {Zhao}, {Shi}, {Luo}, {Li}, {Wu},
  {Bai}, {Zhang}, {Hou}, {Yuan}, {Li}, \& {Wei}}]{Xiang2015}
{Xiang}, M.~S., {et~al.} 2015, \mnras, 448, 822

\bibitem[{{Yang} {et~al.}(2012){Yang}, {Carlin}, {Liu}, {Zhang}, {Gao}, {Xu},
  {Deng}, {Newberg}, {L{\'e}pine}, {Hou}, {Liu}, {Christlieb}, {Zhang}, {Lee},
  {Pan}, {Han}, \& {Wang}}]{Yangf2012}
{Yang}, F., {et~al.} 2012, Res. Astron. Astrophys., 12, 781

\bibitem[{{Yuan} {et~al.}(2015){Yuan}, {Liu}, {Huo}, {Xiang}, {Huang}, {Chen},
  {Zhang}, {Sun}, {Wang}, {Zhang}, {Zhao}, {Luo}, {Shi}, {Li}, {Yuan}, {Dong},
  {Li}, {Hou}, \& {Zhang}}]{Yuan2015}
{Yuan}, H.-B., {et~al.} 2015, \mnras, 448, 855

\bibitem[{{Zacharias} {et~al.}(2013){Zacharias}, {Finch}, {Girard}, {Henden},
  {Bartlett}, {Monet}, \& {Zacharias}}]{Zacharias2013}
{Zacharias}, N., {Finch}, C.~T., {Girard}, T.~M., {Henden}, A., {Bartlett},
  J.~L., {Monet}, D.~G., \& {Zacharias}, M.~I. 2013, \aj, 145, 44

\bibitem[{{Zhang} \& {Jeffery}(2012)}]{Zhang2012}
{Zhang}, X., \& {Jeffery}, C.~S. 2012, \mnras, 419, 452

\bibitem[{{Zhang} {et~al.}(2012){Zhang}, {Carlin}, {Yang}, {Liu}, {Deng},
  {Newberg}, {Zhang}, {L{\'e}pine}, {Xu}, {Gao}, {Christlieb}, {Han}, {Hou},
  {Lee}, {Liu}, {Pan}, \& {Wang}}]{Zhangy2012a}
{Zhang}, Y.-Y., {et~al.} 2012, Res. Astron. Astrophys., 12, 792

\bibitem[{{Zhao} {et~al.}(2012){Zhao}, {Zhao}, {Chu}, {Jing}, \&
  {Deng}}]{Zhao2012}
{Zhao}, G., {Zhao}, Y.-H., {Chu}, Y.-Q., {Jing}, Y.-P., \& {Deng}, L.-C. 2012,
  Res. Astron. Astrophys., 12, 723

\bibitem[{{Zhao}(2014)}]{Zhaoy2014}
{Zhao}, Y. 2014, in Society of Photo-Optical Instrumentation Engineers (SPIE)
  Conference Series, Vol. 9145, Society of Photo-Optical Instrumentation
  Engineers (SPIE) Conference Series, 17

\end{thebibliography}

\begin{deluxetable}{lrrrrrrrrr}
\tabletypesize{\scriptsize}
\tablecaption{Parameters of 44 composite spectra hot subdwarf stars observed in LAMOST DR1.\label{tbl1-1}}
\tablewidth{0pt}
\tablehead{\colhead{LAMOST}& \colhead{Name}& \colhead{Type}& \colhead{$u$}& \colhead{$g$}& \colhead{$r$}& \colhead{$V$}& \colhead{$V-J$}& \colhead{$J-H$}\\
\colhead{}& \colhead{}& \colhead{}& \colhead{mag}& \colhead{mag}& \colhead{mag}& \colhead{mag}& \colhead{mag} & \colhead{mag}}
\startdata
J001227.76+035431.7 &PG0009+036              &sdB+MS &  --  &13.01 &13.28 &13.134 &$-$0.221 &$-$0.076\\
J011929.04+242531.2 &PG0116+242              &sdB+MS &  --  &11.93 &11.61 &11.716 &1.062  & 0.255\\
J012952.60+320209.6 &PG0127+3146             &sdB+MS &  --  &14.13 &14.52 &14.423 &0.003  & 0.128\\
J015055.13+025239.5 &     --                 &sdB+MS &16.17 &16.29 &16.49 &16.354 &0.331  & 0.421\\
J020001.63+140942.5 &2MassJ02000162+1409419  &sdB+MS &  --  &  --  &12.11 &12.970 &1.038  & 0.255\\
J030342.80+012854.8 &KUV03011+0117           &sdB+MS &17.82 &16.75 &16.49 &16.534 &0.917  & 0.182\\
J034252.43+045305.7 &--                      &sdB+MS &14.02 &  --  &14.13 &   --  &  --   & 0.359\\
J042634.61+165526.2 &--                      &sdB+MS &  --  &14.00 &13.86 &13.945 &1.154  & 0.286\\
J071007.73+342453.0 &BD+34 1543              &sdB+MS &  --  &  --  & 9.95 &10.156 &0.671  & 0.159\\
J073712.27+264224.7 &SDSSJ073712.27+264224.7 &sdB+MS &  --  &15.00 &15.21 &15.147 &0.356  & 0.265\\
J081406.83+201901.1 &--                      &sdB+MS &15.93 &15.74 &15.91 &15.563 &0.115  & 0.228\\
J082517.99+113106.3 &--                      &sdB+MS &14.77 &14.69 &14.74 &14.376 &0.364  & 0.354\\
J084408.20+310211.0 &PG0841+312              &sdB+MS &  --  &14.46 &14.70 &14.599 &0.411  & 0.324\\
J093541.33+162110.9 &PG0932+166              &sdB+MS &14.61 &14.67 &14.97 &14.816 &0.476  & 0.443\\
J101317.96+362507.3 &KUV10104+3640           &sdB+MS &16.03 &15.14 &14.94 &15.000 &0.855  & 0.235\\
J101640.84$-$010900.5&SDSSJ10640.84$-$010900.5&sdB+MS &16.45 &16.29 &16.42 &   --  &   --  & 0.356\\
J102234.91+460058.7 &SDSSJ102234.91+460058.7&sdB+MS &17.17 &16.70 &16.62 &16.932 & 0.814 &$-$0.027\\
J103638.93+195202.2 &PG1033+201             &sdB+MS &  --  &15.40 &15.80 &15.637 & 0.180 & 0.299\\
J110403.08+523712.6 &PG1101+529             &sdB+MS &15.19 &14.86 &14.86 &14.878 & 0.574 & 0.129\\
J111436.51+334027.0 &FBS1111+339            &sdB+MS &  --  &12.52 &12.35 &12.400 & 1.128 & 0.486\\
J112213.10+142621.7 &PG1119+147             &sdB+MS &16.22 &16.33 &16.62 &16.262 & 0.414 & 0.413\\
J120341.17+253111.4 &PG1201+258             &sdB+dM &14.77 &14.98 &15.44 &15.164 & 0.037 & 0.308\\
J121238.56+424002.2 &PG1210+429             &sdB+MS &15.09 &14.98 &15.04 &14.960 & 0.484 & 0.270\\
J121735.90+375824.9 &FBS1215+382            &sdB+MS &15.84 &15.78 &15.99 &15.693 & 0.382 & 0.235\\
J123451.01+494720.2 &PG1232+501             &sdB+MS &14.03 &16.24 &14.13 &13.955 & 0.249 & 0.140\\
J125004.42+550602.1 &GD 319                 &sdB+MS &  --  &12.26 &12.28 &12.259 & 0.711 & 0.461\\
J130013.83$-$024952.5&PG1257$-$026          &sdB+MS &  --  &  --  &13.62 &14.036 & 0.506 & 0.211\\
J130025.53+004530.1 &PG1257+010             &sdB+MS &15.85 &15.98 &16.19 &15.847 & 0.438 & 0.402\\
J131248.79+174101.6 &PG1310+179             &sdB+MS &15.24 &15.48 &15.78 &15.37  & 0.192 & 0.290\\
J132917.48+542027.5 &PG1327+546             &sdB+MS &  --  &14.91 &14.54 &14.676 & 0.790 & 0.256\\
J140117.20+273841.7 &PG1359+279             &sdB+MS &16.26 &16.01 &16.11 &16.201 & 0.641 & 0.212\\
J140203.86+072539.1 &PG1359+077             &sdB+MS &15.96 &16.15 &16.49 &16.099 & 0.442 & 0.239\\
J153203.25+425745.8 &PG1530+431             &sdB+MS &15.18 &15.23 &15.41 &15.240 & 0.306 & 0.279\\
J154124.97+290130.1 &PG1539+292             &sdB+MS &14.90 &14.98 &16.36 &14.630 & 0.444 & 0.191\\
J154210.88+015557.2 &       --              &sdB+MS &16.33 &16.21 &16.24 &15.947 & 0.713 & 0.235\\
J163201.35+075940.0 &PG1629+081             &sdB+MS &  --  &12.61 &12.90 &12.762 &$-$0.074 & 0.225\\
J170716.53+275410.4 &      --               &sdB+MS &17.10 &16.98 &17.11 &16.568 & 0.461 & 0.436\\
J170959.18+405450.1 &PG1708+409             &sdB+MS &  --  &15.04 &15.30 &15.222 & 0.255 & 0.279\\
J172627.93+370919.4 &FBS1724+372            &sdB+MS &  --  &  --  &11.86 &13.363 & 0.572 & 0.141\\
J175403.69+534135.6 &2MassJ17540354+5341359 &sdB+MS &15.41 &15.36 &15.41 &15.298 & 0.525 & 0.278\\
J221830.58+184808.8 &HS2216+1833            &sdB+MS &  --  &13.84 &13.92 &16.095 &$-$0.429 & 0.099\\
J230233.84+260257.9 &2MassJ23023384+2602579 &sdB+MS &15.33 &14.80 &14.61 &14.684 & 0.872 & 0.211\\
J232105.80+241039.0 &PG2318+239             &sdB+MS &  --  &  --  &14.00 &13.656 &$-$0.218 &$-$0.090\\
J232147.42+251650.8 &Balloon93738002        &sdB+MS &  --  &13.48 &13.47 &13.470 & 0.643 & 0.242\\
\enddata
\end{deluxetable}

\begin{deluxetable}{lrrrrrrrrrrrrr}
\setlength{\tabcolsep}{0.04in}
\rotate
\tabletypesize{\scriptsize}
\tablecaption{Atmospheric parameters of 122 non-composite spectra hot subdwarf stars observed in LAMOST DR1. $n(\rm{He})$$+n(\rm{H})=1$ and $\log(y)=\log(n(\rm{He})/$$(1-n(\rm{He})))$.  \label{tbl2-2}}
\tablewidth{0pt}
\tablehead{\colhead{LAMOST}& \colhead{Name}& \colhead{$T_{\rm{eff}}$}& \colhead{$\log$(g)}& \colhead{$\log(y)$}& \colhead{Type$^{b}$}& \colhead{$u$}& \colhead{$g$}& \colhead{$r$}& \colhead{$V$}& \colhead{$V-J$}& \colhead{$J-H$} & \colhead{SNR}  & \colhead{Ref} \\
\colhead{}& \colhead{}&\colhead{($\rm{K}$)}&\colhead{($\rm{cm\,s}^{-2}$)} &\colhead{} & \colhead{}& \colhead{mag}& \colhead{mag}& \colhead{mag}& \colhead{mag}& \colhead{mag} & \colhead{mag} &\colhead{} & \colhead{}  }
\startdata
  J000106.72+110036.3   &     PG2358+107            & $27100\pm430$     & $5.541\pm0.067$ & $-2.724\pm0.109$ & sdB    &   --    & 13.46 & 13.79 & 13.593 & $-$0.470 & $-$0.128 & 57.4 &   \\
  J002747.80+344026.5   &     HS0025+3423           & $32320\pm-1430$   & $5.746\pm0.206$ & $-0.917\pm0.156$ & He-sdB &   --    & 15.54 & 16.06 & 15.849 & $-$0.398 & $-$0.003 & 17.4 &1  \\
  J005824.66+015435.0   &  PG0055+016               & $32350\pm450$     & $5.713\pm0.118$ & $-1.925\pm0.114$ & sdB    &   14.52 & 14.85 & 15.35 & 15.099 & $-$0.736 & $-$0.120 & 41.9 &1  \\
  J010421.67+041337.0   &     PG0101+039            & $27160\pm520$     & $5.459\pm0.05$  & $-2.771\pm0.096$ & sdB    &   --    & 11.86 & 12.23 & 11.982 & $-$0.667 & $-$0.107 &100.8 &7  \\
  J011857.19$-$002545.5 & SDSSJ011857.20$-$002546.5 & $28650\pm1230$    & $5.592\pm0.057$ & $-3.031\pm0.295$ & sdB    &   14.49 & 14.60 & 15.07 & 14.804 & $-$0.380 & $-$0.078 &48.4  &   \\
  J011928.87+490109.3   & GALEXJ011928.88+490109.39 & $42660\pm-780$    & $5.841\pm0.211$ & $0.295\pm0.315$  & He-sdO &   --    &  --   & 13.54 & 13.432 & $-$0.372 & $-$0.257 &21.3  &6  \\
  J024734.99+364550.3   &    KUV02445+3633          & $42590\pm340$     & $5.666\pm0.101$ & $1.782\pm1.058$  & He-sdO &   --    &  --   & 13.65 & 13.013 & $-$0.583 & $-$0.115 &108.4 &   \\
  J030128.00+301536.6   & HS0258+3003               & $55880\pm4670$    & $6.252\pm0.182$ & $-2.014\pm0.442$ & sdO    &   --    &  --   & 15.43 & 14.993 & $-$0.406 & $-$0.030 & 31.6 &  \\
  J032138.67+053839.9   & PG0319+055                & $31690\pm240$     & $5.728\pm0.068$ & $-2.034\pm0.095$ & sdB    &   14.86 & 14.88 & 15.14 & 15.048 & $-$0.100 & $-$0.254 & 62.7 &  \\
  J034208.81+090220.7   &--                         & $40420\pm2450$    & $5.275\pm0.222$ & $-2.827\pm0.814$ & sdO    &   15.81 & 15.88 &16.06 & 15.913  & 0.093    & $-$0.271 & 20.3 &  \\
  J035952.18+014208.5   & HS0357+0133               &$28630\pm720$      & $5.742\pm0.115$ & $-2.392\pm0.213$ & sdB    &   --    &  --   & 15.12 & 14.936 & $-$0.038 & $-$0.187 & 41.6 &1  \\
  J062407.08+294721.6   & KUV06209+2949             &$15120\pm420$      & $4.897\pm0.094$ & $-1.971\pm0.406$ & BHB    &   --    & 16.68 & 16.75 & 16.737 & 0.058    & $-$0.411 & 14.3 &  \\
  J065251.96+290023.7   & SDSSJ065251.84+290023.2   &$32110\pm580$      & $5.699\pm0.127$ & $-1.912\pm0.122$ & sdB    &   14.32 & 14.57 & 15.03 & 14.820 & $-$0.559 & $-$0.028 & 35.2 &  \\
  J065658.95+284458.3   & SDSSJ065658.94+284457.6   &$29660\pm1100$     & $5.551\pm0.162$ & $-3.203\pm0.754$ & sdB    &   16.42 &  --   & 16.97 & 16.912 & --       & --       & 21.4 &  \\
  J070147.91+283405.3   & KUV06586+283              &$26070\pm930$      & $5.485\pm0.129$ & $-1.880\pm0.147$ & sdB    &   14.72 & 14.73 & 15.11 & 14.834 & $-$0.542 & 0.025    & 26.0 &  \\
  J072351.47+301916.5   & SDSSJ072351.47+301916.5   &$31820\pm1240$     & $5.743\pm0.233$ & $-1.496\pm0.24$  & sdB    &   --    & 14.73 & 15.27 & 15.043 & $-$0.558 & $-$0.126 & 20.9 &4  \\
  J074613.16+333307.5   & SDSSJ074613.16+333307.7   & $47270\pm2530$    & $5.773\pm0.245$ & $0.500\pm0.423$  & He-sdO &   --    & 15.61 & 16.09 & 15.944 & $-$0.424 & $-$0.258 & 20.2 &4  \\
  J080628.09+323059.4   & 2MJ080628.09+323059.4     & $32450\pm580$     & $5.924\pm0.115$ & $-1.357\pm0.097$ & sdB    &   --    & 15.08 & 15.60 & 15.472 & $-$0.549 & $-$0.098 & 39.4 &  \\
  J080656.76+152718.1   & 2MJ08065668+1527200       & $28960\pm1800$    & $5.302\pm0.321$ & $-2.726>$        & sdB    &   14.41 & 14.54 & 15.03 & 14.775 & $-$0.658 & 0.377    & 14.5 &   \\
  J081204.87+135205.1   & KUV06586+2838             & $24010\pm1940$    & $4.666\pm0.225$ & $-3.101\pm0.873$ & BHB    &   16.94 & --    & 17.54 & 17.531 & --       &  --      & 15.7 &   \\
  J081351.59+110136.3   & --                        & $20120\pm3590$    & $4.576\pm0.145$ & $-3.228\pm0.802$ & BHB    &   14.86 & 15.12 & 15.61 & 15.47  & $-$0.566 & $-$0.754 & 25.2 &  \\
  J082226.26+394119.0   & KUV08191+3951             & $31160\pm980$     & $5.840\pm0.226$ & $-2.244\pm0.29$  & sdB    &   16.79 & --    & 17.43 & 17.126 & --       &  --      & 20.5 &   \\
  J082802.03+404008.8   & WD0824+408                &$59350\pm18890$    & $4.926\pm0.401$ & $-1.800\pm1.241$ & sdO    &   17.59 & --    & 18.52 & 17.95  & --       &  --      & 8.4  &4   \\
  J083603.96+155215.4   & SDSSJ083603.98+155216.4   & $27100\pm640$     & $5.419\pm0.069$ & $-2.451\pm0.174$ & sdB    &   15.07 & 15.18 & 15.64 & 15.406 & $-$0.567 & 0.04     & 57.4 &   \\
  J085323.65+164935.2   & PG0850+170                & $27090\pm740$     & $5.398\pm0.079$ & $-2.78\pm0.183$  & sdB    &   --    & 13.73 & 14.18 & 13.998 & $-$0.548 & $-$0.001 & 63.3 &  \\
  J085649.26+170114.6   & GALEXJ085649.30+170115.0  & $29360\pm230$     & $5.477\pm0.064$ & $-3.101\pm0.199$ & sdB    &   --    & 12.61 & 13.07 & 12.84  & $-$0.586 & $-$0.079 &102.9 &6  \\
  J085902.64+115627.7   & PG0856+121                & $25010\pm1040$    & $5.525\pm0.106$ & $-3.162>$        & sdB    &   --    & 13.3  & 13.79 & 13.48  & $-$0.473 & $-$0.095 & 39.2 &7  \\
  J090447.76+313252.7   & PG0901+309                & $38400\pm900$     & $5.685\pm0.305$ & $-0.57\pm0.145$  & He-sdB &   14.55 & 14.9  & 15.40 & 15.172 & $-$0.591 & $-$0.099 & 25.5 & 3  \\
  J091025.43+120827.0   & PG0907+123                & $27560\pm440$     & $5.324\pm0.064$ & $-2.836\pm0.136$ & sdB    &  --     & 13.75 & 14.14 & 13.916 & $-$0.558 & $-$0.192 & 89.7 & 7  \\
  J091207.29+161320.4   & PG0909+164                & $31670\pm950$     & $4.73\pm0.291$  & $-2.245\pm0.494$ & sdB    &  --     & 13.57 & 14.09 & 13.851 & $-$0.634 & $-$0.200  &16.1  & 3,8  \\
  J091251.66+272031.4   & PG0909+276                & $37560\pm310$     & $5.970\pm0.076$ & $-0.908\pm0.042$ & He-sdB  &  --    & 12.06 & 12.49 & 12.276 & $-$0.584 & $-$0.102  &142.8 & 8  \\
  J091408.68+035804.0   & PG0911+042                & $27980\pm1370$    & $5.505\pm0.268$ & $-2.993>$        & sdB     &  --    &  --   & 15.10 & 15.486 & $-$0.514 & 0.207     & 15.0 &   \\
  J092128.21+024602.3   & PG0918+029                & $31460\pm510$     & $5.788\pm0.116$ & $-2.531\pm0.144$ & sdB     &  --    &  --   & 13.75 & 13.303 & $-$0.646 & $-$0.208  &69.3  &    \\
  J092239.83+270225.4   & PG0919+273                & $33230\pm240$     & $5.997\pm0.041$ & $-2.395\pm0.129$ & sdB     &  --    & 12.41 & 12.90 & 12.658 & $-$0.645 & $-$0.117  &72.3  &    \\
  J092308.30+024209.9   & PG0920+029                & $29980\pm910$     & $5.472\pm0.174$ & $-3.515>$        & sdB     &  --    &  --   & 14.03 & 14.352 & $-$0.711 & $-$0.180  & 29.1 &   \\
  J092313.41+292657.5   & PG0920+297                & $30810\pm1430$    & $5.993\pm0.221$ & $-1.284\pm0.181$ & sdB     &  14.19 & 14.43 & 15.79 & 14.729 & $-$0.683 & $-$0.201  & 22.3 &   \\
  J092830.55+561811.7   & PG0924+565                &$58780\pm2740$    & $5.270\pm0.175$ & $-1.080\pm0.257$  & sdO     &  15.49 & 15.95 & 16.45 & 15.911 & $-$0.723 & 0.659     & 25.3 & 4   \\
  J093015.51+305034.6   & PG0927+311                & $28140\pm490$     & $5.809\pm0.068$ & $-2.679\pm0.124$ & sdB     &  14.49 & 14.65 & 15.11 & 14.956 & $-$0.475 & $-$0.152  & 49.2 &    \\
  J093512.15+311000.4   & PG0932+314                & $33440\pm670$     & $5.841\pm0.139$ & $-1.595\pm0.133$ & sdB     &  15.08 & 15.35 & 15.87 & 15.634 & $-$0.765 & $-$0.070  & 38.6 &4  \\
  J093716.27+182511.2   & PG0934+186                & $34970\pm1060$    & $5.575\pm0.139$ & $-2.525>$        & sdB     &  --    & 12.86 & 13.37 & 13.131 & $-$0.628 & $-$0.213  & 107.3&   \\
  J093820.35+550550.0   & PG0934+553                & $44340\pm350$     & $5.373\pm0.056$ & $-0.432\pm0.231$ & He-sdO  &  --    & 11.87 & 12.21 & 12.019 & $-$0.284 & 0.148     &148.8 &    \\
  J094623.10+040456.0   & PG0943+043                & $37110\pm1020$    & $5.771\pm0.200$ & $-1.453\pm0.204$ & sdB     &  15.23 & 15.49 & 15.97 & 15.735 & $-$0.529 & 0.254     & 15.9 & 4  \\
  J094729.40+271627.0   & PG0944+275                & $28320\pm1720$    & $5.893\pm0.228$ & $-2.262>$        & sdB     &  16.21 & --    & 16.89 & 16.706 & --       &  --       & 11.0 &   \\
  J095058.04+182618.5   & PG0948+187                & $35340\pm970$     & $5.847\pm0.192$ & $-1.844\pm0.202$ & sdB     &  15.93 & 16.21 & 16.72 & 16.046 & $-$0.256 & 0.429     & 36.4 &4   \\
  J095101.34+034757.3   & PG0948+041                & $31000\pm1360$    & $5.531\pm0.206$ & $-3.543\pm1.122$ & sdB     &  15.59 & 15.73 & 16.12 & 15.817 & $-$0.155 & 0.206     & 30.9 &    \\
  J095932.25+361825.8   & CBS 115                   & $27180\pm230$     & $5.224\pm0.031$ & $-2.694\pm0.078$ & sdB     &  --    & 12.65 & 13.12 & 12.905 & $-$0.516 & $-$0.140  &119.9 &    \\
  J095952.08+033032.6   & PG0957+037                & $36640\pm1630$    & $5.152\pm0.145$ & $-3.505\pm1.116$ & sdB     &  14.89 & 15.16 & 15.68 & 15.448 & $-$0.390 & 0.069     & 32.0 &   \\
  J100154.98+301805.6   & SDSSJ100154.98+301805.6   & $23290\pm2160$    & $4.665\pm0.314$ & $-2.141>$        & BHB     &  15.84 & --    & 16.44 & 16.148 & --       &  --       &  6.2 &    \\
  J100354.27+403418.1   & PG1000+408                & $40990\pm1030$    & $5.220\pm0.070$ & $-2.884\pm0.290$ & sdO     &  --    & 12.97 & 13.57 & 13.289 & $-$0.689 & $-$0.266  & 83.9 &   \\
  J101342.12+260620.0   & SDSSJ101342.12+260619.9   & $47160\pm5190$    & $5.807\pm0.310$ & $-1.701\pm0.425$ & sdO     &  16.32 & --    & 17.24 & 16.638 & --       &  --       &24.7  &4   \\
  J101420.74$-$025228.1$^{a}$&     --                & $50660\pm3280$    & $5.702\pm0.378$ & $0.579\pm1.168$  & He-sdO  &  15.56 & 15.96 & 16.51 & 16.334 & $-$0.417& $-$0.120  & 20.8 &   \\
  J102029.80+425021.9   & PG1017+431                & $40230\pm1340$    & $5.114\pm0.095$ & $-2.783\pm0.252$ & sdO     &  14.6  & 14.95 & 15.53 & 15.309 & $-$0.711 & $-$0.099  & 68.4 & 3  \\
  J102120.45+444636.9   & SDSSJ102120.44+444636.9   & $48250\pm3920$    & $5.746\pm0.592$ & $0.721\pm0.979$  & He-sdO  &  17.31 & --    & 18.29 & 18.241 & --       &  --       & 9.6  & 4  \\
  J103516.57+402114.4   & PG1032+406                & $31920\pm210$     & $5.840\pm0.063$ & $-2.253\pm0.056$ & sdB     &  --    & 11.31 & 11.72 & 11.474 & $-$0.692 & $-$0.109  &168.8 &    \\
  J104123.24+504419.9   & PG1038+510                & $51720\pm1690$    & $5.884\pm0.199$ & $0.588\pm0.569$  & He-sdO  &  14.33 & 14.73 & 15.27 & 15.008 & $-$0.645 & $-$0.224  & 28.2 &    \\
  J105418.52+494959.7   & PG1051+501                & $34120\pm300$     & $5.130\pm0.053$ & $-1.458\pm0.061$ & sdB     &  --    & 13.12 & 13.64 & 13.381 & $-$0.715 & $-$0.085  & 93.6 &   \\
  J105428.85+010514.8   & SDSSJ105428.85+010514.7   & $27600\pm2150$    & $5.853\pm0.270$ & $-2.978\pm0.787$ & sdB     &  16.56 & --    & 17.24 &  --     &  --      &  --      & 19.9 &   \\
  J111904.87+295153.5   & PG1116+301                & $31580\pm1030$    & $6.165\pm0.249$ & $-2.359\pm0.333$ & sdB     &  13.85 & 14.04 & 14.56 & 14.369 & $-$0.576 & 0.31      & 15.4 &    \\
  J112637.06+115959.8   & PG1124+123                & $27910\pm1090$    & $5.144\pm0.216$ & $-3.572\pm0.790$ & sdB     &  15.67 & 15.78 & 16.3  & 15.695 & $-$0.463 & 0.111     & 29.1 &   \\
  J112829.30+291504.7   & PG1125+295                & $49710\pm1530$    & $5.847\pm0.249$ & $-2.073\pm0.520$ & sdO     &  14.46 & 14.86 & 15.38 & 15.187 & $-$0.742 & $-$0.434  &18.2 & 4  \\
  J113003.83+013738.1   & PG1127+019                & $43650\pm680$     & $5.944\pm0.191$ & $1.947\pm1.158$  & He-sdO  &  13.19 & 13.57 & 14.09 & 13.853 & $-$0.645 & $-$0.253  &31.1 & 7  \\
  J113257.47+050648.8   & PG1130+054                & $30630\pm520$     & $5.960\pm0.132$ & $-3.211>$        & sdB     &  14.65 & 14.71 & 15.18 & 14.885 & $-$0.440 & $-$0.338  & 35.0 &   \\
  J113340.54+560624.2   & PG1130+564                & $31900\pm1160$    & $5.060\pm0.244$ & $-2.696\pm0.493$ & sdB     &  14.82 & 15.03 & 15.41 & 15.275 & $-$0.046 & 0.157     & 24.9 &    \\
  J113942.01+464349.4   & PG1137+470                & $30700\pm460$     & $5.533\pm0.091$ & $-3.982>$        & sdB     &  --    & 15.29 & 15.78 & 15.595 & $-$0.569 & 0.348     & 55.8 &   \\
  J115435.80+582956.7   & SBSS1152+587              & $35510\pm1980$    & $5.931\pm0.197$ & $-0.503\pm0.136$ & He-sdB  &  17.11 & --    & 17.78 & 17.803 & --       &  --       &  8.0 &  \\
  J120624.41+570936.3   & PG1203+574                & $35080\pm880$     & $5.805\pm0.127$ & $-1.866\pm0.147$ & sdB     &  14.24 & 14.52 & 15.08 & 14.894 & $-$0.682 & $-$0.115  & 40.7 &    \\
  J123551.14+422239.7   & PG1233+427                & $26590\pm490$     & $5.479\pm0.049$ & $-2.543\pm0.129$ & sdB     &  --    & 11.26 & 12.35 & 12.046 & $-$0.551 & $-$0.119  & 68.8 &   \\
  J123652.66+501513.5   & PG1234+505                & $42190\pm1200$    & $5.372\pm0.068$ & $-2.220\pm0.225$ & sdO     &  --    &  14.4 & 14.96 & 14.68  & $-$0.768 & $-$0.234  & 45.7 &    \\
  J124201.73+434023.3   & PG1239+439                & $37400\pm1720$    & $5.669\pm0.225$ & $-0.229\pm0.198$ & He-sdB  &  16.66 & --    & 17.47 & 16.819 & --       &  --       & 12.9 & 4  \\
  J124451.20+435252.5   & PG1242+442                & $29760\pm1790$    & $5.496\pm0.259$ & $-2.823>$        & sdB     &  16.11 & --    & 16.72 & 16.498 & --       &  --       & 19.4 &   \\
  J125050.26+161003.1   & PG1248+164                & $24830\pm1450$    & $5.453\pm0.202$ & $-2.517\pm0.255$ & sdB     &  --    & 14.23 & 14.75 & 14.46  & $-$0.577 & 0.024     & 24.4 &    \\
  J125229.60$-$030129.6 & PG1249$-$028              & $30780\pm480$     & $5.694\pm0.128$ & $-4.500\pm1.336$ & sdB     &  15.46 & 15.71 & 16.22 & 15.618 & $-$0.541 & $-$1.430  & 34.4 &   \\
  J125318.50+300629.3   & PG1250+304                & $32550\pm510$     & $5.809\pm0.098$ & $-2.278\pm0.165$ & sdB     &  15.61 & 15.91 & 16.45 & 15.939 & $-$0.448 & $-$1.160  & 86.9 &   \\
  J125627.45+274230.6   & PG1254+279                & $25050\pm4010$    & $5.545\pm0.377$ & $-2.522>$        & sdB     &  15.52 & 15.59 & 16.07 & 15.861 & $-$0.646 & 0.146     &  12.1 &  \\
  J125926.03+272122.7   & PG1257+276                & $18430\pm390$     & $4.883\pm0.073$ & $-1.705\pm0.107$ & BHB     &  --    & 15.09 & 15.56 & 15.43  & $-$0.263 & $-$0.512  & 43.5 &   \\
  J130346.61+264630.6   & PG1301+270                & $49400\pm1280$    & $6.538\pm0.175$ & $-0.097\pm0.260$ & He-sdO  &  --    & 15.34 & 15.88 & 15.71  & $-$0.505 & 0.112     & 23.8 &   \\
  J130448.68+280729.9   & PG1302+284                & $34580\pm1070$    & $5.758\pm0.160$ & $-2.841>$        & sdB     &  14.94 & 15.23 & 15.77 & 15.543 & $-$0.631 & $-$0.033  & 21.0 &   \\
  J130615.56+485019.7   & PG1304+491                & $32430\pm290$     & $5.682\pm0.040$ & $-1.767\pm0.054$ & sdB     &  13.21 & 15.51 & 14.07 & 13.725 & $-$0.689 & $-$0.084  & 29.7 &   \\
  J132044.38+055901.3   & PG1318+062                & $44560\pm1040$    & $5.791\pm0.206$ & $1.084\pm0.755$  & He-sdO  &  14.07 & 14.5  & 14.98 & 14.786 & $-$0.583 & $-$0.134  & 23.9 &   \\
  J132434.93+281802.3   & PG1322+286                & $32500\pm1710$    & $5.818\pm0.361$ & $-2.348>$        & sdB     &  14.71 & 14.93 & 15.47 & 15.179 & $-$0.736 & $-$0.189  &  14.5 &  \\
  J133153.55+154117.5   & PG1329+159                & $29480\pm950$     & $5.560\pm0.173$ & $-2.767>$        & sdB     &  --    & 13.28 & 13.72 & 13.507 & $-$0.528 & $-$0.148  & 30.1 &   \\
  J133338.07+584933.7   & PG1331+591                & $33400\pm590$     & $5.136\pm0.081$ & $-0.988\pm0.066$ & He-sdB  &  14.48 & 14.71 & 15.19 & 14.981 & $-$0.356 & 0.133     & 42.5 &    \\
  J134008.83+475151.9   & PG1338+481                & $28360\pm300$     & $5.501\pm0.049$ & $-2.823\pm0.133$ & sdB     &  --    & 13.10 & 13.79 & 13.588 & $-$0.578 & $-$0.152  &100.2 &   \\
  J134131.48+045446.7   & PG1339+052                & $61370\pm9290$    & $6.304\pm0.218$ & $-1.607\pm0.377$ & sdO     &  15.87 & 16.28 & 16.85 & 16.152 & $-$0.417 & 0.092     & 25.2 &  \\
  J135015.85+602438.4   & PG1348+607                & $54360\pm1980$    & $5.448\pm0.251$ & $-0.062\pm0.575$ & He-sdO  &  15.59 & --    & 16.61 & 16.66  & --       &  --       & 22.8 &  3 \\
  J135153.11$-$012946.6 & PG1349$-$012              & $30970\pm920$     & $5.671\pm0.187$ & $-2.631>$        & sdB     &  15.31 & 15.45 & 15.90 & 15.964 & $-$0.228 & $-$0.171  & 33.7 &   \\
  J135824.65+065135.3   & PG1355+071                & $24310\pm910$     & $5.705\pm0.124$ & $-2.882\pm0.418$ & sdB     &    --  & 14.16 & 14.59 & 14.346 & $-$0.628 & $-$0.017  & 11.6 &    \\
  J140545.25+014419.0   & PG1403+019                & $30300\pm990$     & $5.848\pm0.165$ & $-2.264\pm0.311$ & sdB     &  15.9  & 15.91 & 16.33 & 15.787 & $-$0.519 & $-$0.089  & 29.6 & 5  \\
  J141702.82+485725.8   & PG1415+492                & $37690\pm1790$    & $5.233\pm0.248$ & $2.459\pm1.151$  & He-sdB  &  13.76 & 14.07 & 14.56 & 14.299 & $-$0.667 & $-$0.065  & 76.5 & 7  \\
  J141736.40$-$043429.0 & PG1415$-$043              & $38030\pm540$     & $5.884\pm0.120$ & $-1.568\pm0.117$ & sdB     &  --    & 13.52 & 13.96 & 13.724 & $-$0.653 & $-$0.141  & 53.4 &  \\
  J143729.14$-$021506.0 &     --                    & $35870\pm1180$    & $5.696\pm0.125$ & $0.000\pm0.135$  & He-sdB  &  15.65 & 15.96 & 16.44 & 15.742 & $-$0.841 & 0.205     & 30.7 &    \\
  J144052.82$-$030852.6 & PG1438$-$029              & $29280\pm240$     & $5.405\pm0.057$ & $-2.893\pm0.145$ & sdB     &  --    & 13.60 & 14.02 & 13.792 & $-$0.376 & $-$0.072  & 104.2&   \\
  J144227.47$-$013245.9 & PG1439$-$013              & $43080\pm2990$    & $4.922\pm0.157$ & $-2.696\pm0.539$ & sdO     &  --    & 13.64 & 14.11 & 13.873 & $-$0.633 & $-$0.186  & 29.2 &    \\
  J144708.27+072349.5   & PG1444+076                & $50640\pm860$     & $5.760\pm0.102$ & $1.237\pm1.075$  & He-sdO  &  --    & 14.42 & 14.94 & 14.717 & $-$0.667 & $-$0.119  & 62.5 &   \\
  J144933.64+244336.2   & PG1447+249                & $36590\pm950$     & $5.498\pm0.156$ & $-1.749\pm0.288$ & sdB     &  --    & 15.45 & 15.96 & 15.799 & $-$0.572 & $-$0.251  & 17.4 &   \\
  J151030.69$-$014345.8 & PG1507$-$015              & $45800\pm1210$    & $6.251\pm0.256$ & $0.297\pm0.460$  & He-sdO  &  15.91 & 16.25 & 16.71 &  --     &  --     &  $-$0.403 & 15.8 & 4  \\
  J153329.95+520648.7   & PG1532+523                & $31510\pm470$     & $5.894\pm0.112$ & $-2.401\pm0.172$ & sdB     &  13.45 & 16.10 & 14.23 & 14.007 & $-$0.683 & $-$0.181  & 48.9 & 3   \\
  J154039.03+395549.0   & PG1538+401                & $33800\pm1510$   & $5.906\pm0.185$  & $2.756>$         & sdB     &  --    & 12.80 & 13.46 & 13.216 & $-$0.666 & $-$0.120  &106.7 & 3   \\
  J154611.68+483837.2   & PG1544+488                & $40030\pm1560$   & $6.413\pm0.242$  & $4.717\pm2.177$  & He-sdO  &  --    & 12.48 & 13.04 & 12.792 & $-$0.668 & $-$0.101  & 19.8 & 3  \\
  J154720.93+055937.7   & SDSSJ154720.93+055937.00  & $28570\pm1240$   & $5.462\pm0.134$  & $-1.816\pm0.375$ & sdB     &  16.64 & 16.42 & 16.77 & 16.361 & $-$0.265 & 0.245     & 15.4 &    \\
  J154837.17+042126.9   & PG1546+045                & $32210\pm1150$   & $5.709\pm0.228$  & $-2.483\pm0.471$ & sdB     &  --    & 15.31 & 15.74 & 15.549 & $-$0.316 & $-$0.010  & 32.7 &    \\
  J155144.87+002948.8   & PG1549+006                & $33610\pm1660$   & $5.659\pm0.285$  & $-1.931\pm0.252$ & sdB     &  --    & 14.97 & 15.43 & 15.211 & $-$0.355 & $-$0.105  & 19.0 & 3  \\
  J155537.94+270648.6   & PG1553+273                & $20810\pm310$    & $4.809\pm0.055$  & $-2.810\pm0.195$ & BHB     &  13.44 & 13.39 & 13.77 & 13.53  & $-$0.461 & $-$0.042  & 61.2 &   \\
  J160112.12+531151.9   & PG1559+533                & $31410\pm520$    & $5.690\pm0.145$  & $-2.344\pm0.212$ & sdB     &  --    & 13.9  & 14.54 & 14.288 & $-$0.667 & $-$0.131  & 19.0 & 3   \\
  J160131.27+044027.0   & PG1559+048                & $36520\pm20$     & $5.399\pm0.003$  & $0.394\pm0.235$  & He-sdB  &  --    & 14.22 & 14.66 & 14.455 & $-$0.503 & $-$0.127  & 45.3 & 3  \\
  J160803.68+070428.7   & PG1605+072                & $32550\pm370$    & $5.289\pm0.065$  & $-2.512\pm0.138$ & sdB     &  --    & 12.68 & 13.11 & 12.827 & $-$0.579 & $-$0.079  & 40.1 & 3  \\
  J161200.65+514943.5$^a$ & PG1610+519              & $40270\pm2060$   & $5.580\pm0.162$  & $-2.700\pm0.306$ & sdO     &  13.26 & 13.54 & 13.90 & 13.344 & $-$0.252 & 0.323     & 31.4 &   \\
  J162935.90+003149.1   & PG1627+006                & $18860\pm510$    & $5.770\pm0.100$  & $-2.436\pm0.287$ & sdB     &  --    & 14.76 & 15.13 & 15.000 & $-$0.362 & $-$0.064  & 54.8 & 3  \\
  J163212.26+175318.3   & PG1629+179                & $37440\pm2100$   & $5.716\pm0.192$  & $-3.198\pm0.691$ & sdB     &  15.78 & --    & 16.54 & 15.991 & --       &  --       & 25.6 &    \\
  J164609.24+401725.5   & PG1644+404                & $29990\pm300$    & $5.643\pm0.074$  & $-1.915\pm0.078$ & sdB     &  13.68 & 13.87 & 14.37 & 14.101 & $-$0.632 & $-$0.091  & 76.5 & 7  \\
  J164959.85+533131.7   & PG1648+536                & $32430\pm840$    & $5.515\pm0.141$  & $-7.466\pm0.001$ & sdB     &  --    & 13.68 & 14.34 & 14.092 & $-$0.461 & $-$0.034  & 29.7 & 3  \\
  J170040.65+333747.9   & PG1658+337                & $26540\pm940$    & $5.350\pm0.155$  & $-3.038\pm0.394$ & sdB     &  15.84 & 15.99 & 16.46 & 16.010 & $-$0.377 & 0.227     & 28.2 &  \\
  J170237.68+243522.5   & PG1700+247                & $26420\pm620$    & $5.294\pm0.128$  & $-2.389\pm0.205$ & sdB     &  15.93 & 16.02 & 16.46 & 15.970 & $-$0.589 & 0.808     & 18.2 &   \\
  J170534.62+245326.9   & --                        & $35140\pm1090$   & $5.5744\pm0.157$ & $-1.431\pm0.119$ & sdB     &  16.65 & --    & 17.41 & 16.911 & --       &  --       & 19.4 &   \\
  J171218.75+485835.7   & PG1710+490                & $30050\pm310$    & $5.830\pm0.065$  & $-2.565\pm0.145$ & sdB     &    --  & 12.63 & 13.09 & 12.858 & $-$0.561 & $-$0.194  & 74.9 & 8  \\
  J213526.03$-$065743.4 & PHL48                     & $22470\pm690$    & $4.951\pm0.080$  & $-2.726\pm0.269$ & BHB     &    --  & 13.32 & 13.68 & 13.464 & $-$0.514 & $-$0.101  & 37.0 &   \\
  J220716.49+034219.7   & PG2204+035                & $33130\pm600$    & $5.957\pm0.112$  & $-1.889\pm0.108$ & sdB     &    --  & 14.10 & 14.52 & 14.313 & $-$0.536 & $-$0.193  & 74.9 & 3  \\
  J220800.64+023343.5   & PG2205+023                & $27480\pm980$    & $5.594\pm0.143$  & $-3.350>$        & sdB     &    --  & 13.94 & 14.31 & 14.099 & $-$0.508 & $-$0.096  & 37.1 &  3,8  \\
  J221045.47+014135.6   & HE2208+0126               & $21850\pm1760$   & $5.571\pm0.298$  & $-2.603\pm0.633$ & sdB     &    --  & 15.59 & 16.00 & 13.192 & $-$0.124 & 0.286     & 8.9  &  2   \\
  J222122.56+052458.3   & PG2218+052                & $35950\pm740$    & $5.921\pm0.146$  & $-0.696\pm0.120$ & He-sdB  &    --  & 15.11 & 15.50 & 15.374 & $-$0.420 & $-$0.174  & 32.0 &   \\
  J222159.16+093725.7   & PG2219+094                & $24860\pm370$    & $4.532\pm0.077$  & $-1.333\pm0.047$ & BHB     &    --  & 11.72 & 12.09 & 11.907 & $-$0.308 & $-$0.063  &117.6 &   \\
  J235517.34+182015.3   & PG2352+181                & $48570\pm1340$   & $5.859\pm0.194$  & $1.513\pm1.110$  & He-sdO  &  --    & 13.13 & 13.62 & 13.414 & $-$0.552 & $-$0.150  & 31.6 &  7  \\
\enddata
\tablenotetext{a}{The available spectral range is limited due to poor data quality.}
\tablenotetext{b}{He-deficient sdB = sdB, He-rich sdB = He-sdB, He-deficient sdO = sdO and He-rich sdO = He-sdO.}
\tablerefs{(1) \cite{Edelmann2003}; (2) \cite{Lisker2005}; (3) \cite{Winter2006}; (4) \cite{Hirsch2009}; (5) \cite{Geier2011}; (6) \cite{Nemeth2012}; (7) \cite{Drilling2013}; (8) \cite{Geier2013a}.}

\end{deluxetable}

\end{document}